**Digital N-of-1 Trials and their Application in Experimental Physiology**

Stefan Konigorski [1,2,3,†], Mathias Ried-Larsen[4], Christopher H. Schmid[5]

[1]Hasso Plattner Insitute for Digital Engineering, Potsdam, Germany
[2]University of Potsdam, Digital Engineering Faculty, Potsdam, Germany
[3]Hasso Plattner Institute for Digital Health at Mount Sinai, Icahn School of Medicine at Mount Sinai, New York, NY, USA
[4]The Centre of Inflammation and Metabolism and the Centre for Physical Activity Research, Rigshospitalet, University of Copenhagen, Blegdamsvej 9, 2100 Copenhagen, Denmark.
[5]Department of Biostatistics, School of Public Health, Brown University, Providence, RI, USA

[†]Correspondence: Stefan Konigorski, PhD, Hasso Plattner Institute for Digital Engineering, Prof.-Dr.-Helmert-Str. 1-2, 14482 Potsdam, stefan.konigorski@hpi.de

# New Findings

1. What is the topic of this review?

This review describes the main design and analytic principles behind N-of-1 trials, which use a multi-crossover experimental design in which a single participant receives each of two or more treatments multiple times. Such designs can be used to make statistical inferences about the relative efficacy of the treatments at an individual level. By combining information across different participants, it is also possible to make inferences about the population from which the individuals are drawn.

2. What advances does it highlight?

This review highlights the advantages of using N-of-1 trials, and more specifically digital N-of-1 trials set up through mobile applications, in experimental physiology studies. In particular, it provides an application to the study of the physiology of the acute regulation of response to an exposure with rapid onset and short duration of effect, but can also be applied to long-term effects and adaptive behavior. New developments in models of the responses and ways to describe the effects probabilistically in both a Bayesian and frequentist framework are provided.

# Abstract

Traditionally, studies in experimental physiology have been conducted in small groups of human participants, animal models or cell lines. Identifying optimal study designs that achieve sufficient power for drawing proper statistical inferences to detect group level effects with small sample sizes has been challenging. Moreover, average effects derived from traditional group-level inference do not necessarily apply to individual participants. Here, we introduce N-of-1 trials as an innovative study design that can be used to draw valid statistical inference about the effects of interventions on individual participants and can be aggregated across multiple study participants to provide population-level inferences more efficiently than standard group randomized trials. N-of-1 trials have been used in healthcare settings since the late 1980s, but without large-scale adoption and with few applications in experimental physiology research settings. In this manuscript, we introduce the key components and design features of N-of-1 trials, describe statistical analysis and interpretations of the results, and describe some available digital tools to facilitate their use using examples from experimental physiology.

## 1 Introduction

Historically, clinical, biological and physiological experimental studies have focused on comparing two or more groups of individuals or animals exposed to different interventions (treatments) to determine the relative effects of the interventions on health and function. Ideally, the study subjects are randomized to balance any differences between the groups other than the intervention to which they are assigned. Such randomized experiments directly allow for causal inferences in which the interventions can be said to cause their effects. Though a powerful scientific tool, the typical group randomized experiment in which each participant is treated once only provides inferences about average group effects. It cannot inform about the effect on an individual participant. Another challenge in experimental physiological studies is that often few individuals are available to study leading to low statistical power to detect reasonable size effects. Brennan et al. (2020), Magalhães et al. (2021), Maturana et al. (2021), Melo et al. (2021) and Williams et al. (2019) are examples of randomized trials and pre-post studies in experimental physiology.

Many study designs in which individuals are treated and measured multiple times also cannot estimate individual treatment effects. For instance, longitudinal designs in which individuals are measured repeatedly on one treatment throughout one (Bonafiglia et al., 2019) or multiple (Hecksteden et al., 2015) intervention periods can describe individual trajectories of health outcomes such as cardiorespiratory fitness. But they cannot produce individual-level inferences about intervention effects if each individual is only measured once under one intervention at any time. Two-period crossover trials, in which each participant receives each intervention one time, also cannot estimate individual treatment effects because the outcome is measured only once on each intervention. Examples of two-period crossover trials in experimental physiology are the studies by Thomas et al. (2022), Goltz et al. (2018), Shen et al. (2023) and Yang et al. (2024). Because individual-level treatment effects cannot be estimated in such designs, it is also not possible to validly identify individual treatment responders and non-responders in them (Atkinson, 2019). As pointed out by Lolli et al. (2025), it is common to observe that apparent responders in a first crossover do not respond in the second crossover.

Designs in which crossovers are replicated, however, do provide a means to estimate individual treatment effects and uncover treatment effect heterogeneity. Unless the outcome is measured frequently, however, trials may have low power and large standard errors unless careful design can minimize the effects of biological, contextual or other factors that introduce variability. Chesterton et al. (2021) and Lolli et al. (2025) give examples of replicate (multiple) crossover trials of physiologic outcomes. The astute reader may wonder whether individual treatment effects could be estimated in a two-period crossover study by taking multiple measurements of the outcome in each treatment period, thus creating the replication needed to provide a variance estimate. Lolli et al. (2025) do indeed discuss this issue and indicate that further research is needed. They argue that *"repeated administration of treatments and comparators, including the associated crossover and washout procedures is a fundamentally different process from examining repeated measurements within any given single treatment."* We discuss this issue and potential underlying causes of this observation further in Section 4.

The N-of-1 trial is a replicate (multi-crossover) design whereby each individual receives each treatment (including potentially a control treatment) multiple times, usually in a randomized order. Outcomes are recorded one or more times during each treatment period (e.g., daily for a treatment that is given for one week). Because each individual receives each treatment multiple times, it is possible to obtain statistically valid estimates of individual treatment effects. Although the N-of-1 design can be considered a replicate crossover design, its key characteristic is that it is randomized and carried out on a single person, rather than on a set of people who are randomized as a group.

The N-of-1 design is one type of a class of single-case designs for individuals common in the social sciences, where they take several different forms (Nikles et al., 2015). The randomized multi-crossover N-of-1 design is called a withdrawal/reversal design in the social science literature. Observational N-of-1 studies (Duan, 2022; Kravitz, 2014) are another related class of non-randomized designs, but these are not the focus of this review.

In the medical literature, N-of-1 designs have been used for at least 40 years (Mirza et al., 2017) but have recently become more popular because of the increased recognition of individual variation in treatment response and because of the movement toward personalized health care and scientific self-experimentation by help of digital tools (Daskalova et al., 2021; Konigorski et al., 2022; Selker et al., 2022). Most popularly, N-of-1 trials are recommended for the study of chronic conditions and interventions that have a 1) quick onset, 2) quick washout and 3) no carryover (Nikles et al., 2015; Piccininni et al., 2024). Applications include studies of our own such as in diet (Kaplan et al., 2022), atrial fibrillation (Marcus et al., 2022), chronic pain (Kravitz et al., 2018), fibromyalgia (Zucker et al., 2006), depression (Müller et al., 2023) and more general wellbeing, sleep improvement and stress reduction (Vetter et al., 2024). Recent studies have also applied N-of-1 trials in more complex designs (Gärtner et al., 2023; Meier et al., 2023).

N-of-1 trial designs may be personalized so that individuals design their own trials evaluating interventions and outcomes of interest to them in a manner of their choosing. For example, the PREEMPT trial compared the use of N-of-1 trials to usual care for patients with chronic musculoskeletal pain (Barr et al., 2015; Kravitz et al., 2018). Participants randomized to the N-of-1 arm set up their own trials comparing two treatments of their choice using between two and four intervention periods of one or two weeks on each treatment. They then scored themselves daily on five patient-reported outcomes. Such trials are generally termed 'personalized trials' (Duan et al., 2022; Schmid and Yang, 2022). They enable individuals to choose their own treatments and their favored outcomes to measure. In addition, personalization enables research to expand into new environments including communities and populations underrepresented in traditional research self-experiments that can be conducted scientifically, and to the study of rare or orphan diseases for which etiology, natural history and treatment may be unique to each individual. In personalized trials, but equally in series of identical N-of-1 trials, it is important to be clear about the role of blinding, define estimands explicitly, and be clear whether efficacy or effectiveness of a treatment is studied. For example, if participants are self-selecting their intervention, then inference can only be made about the effect of such a self-selected intervention onto the health outcome of interest, which may differ from the marginal effect of the intervention in the general population. Of note, in many cases, the effect of such self-selected intervention and conditional treatment effects might be of interest to study. Assuming large treatment effect heterogeneity, studies might not aim at estimating average treatment effects in the population but reporting results on the percentage of participants that responded positively to their self-selected treatment. For a more detailed discussion of aspects of blinding and estimands, see sections 3.2 and 4.1.

While the N-of-1 trial fundamentally focuses on providing information to one individual gathered from a single design, patterns from different individuals may also be discerned by analyzing data from multiple N-of-1 trials together. Because the individual responses from each trial from such a series are often available, the combined data are a form of individual participant data meta-analysis (Nikles et al., 2015). Combining trials in this way enables estimating average effects in the population and in subgroups defined by individual design features such as dose or frequency of treatment that vary across individuals. In addition, the combined trial models can improve the estimate of each individual's effect through the principle of borrowing strength from other individuals. Essentially, assuming that individuals arise from a population or subgroup of individuals who can be related through some type of probability structure such as a regression model, inferences about an individual can be improved by combining the information from the individual's own trial with that from trials of other individuals.

We now proceed to explain these concepts in more detail and show their potential use in experimental physiology. The presentation begins in Section 2 by describing an exemplary N-of-1 trial case study of the effect of physical exercise on blood glucose levels that we designed, where we include a discussion of advantages of N-of-1 trials compared to standard randomized trials. This is followed by Section 3 introducing basic features of N-of-1 trials before describing models for data from one and multiple trials in Section 4. We then describe some recent advances in digital technology that facilitate the conduct of N-of-1 trials in Section 5. The paper concludes in Section 6 with discussion and some points to consider when running N-of-1 trials.

## 2 Case study: Effects of physical exercise on blood glucose levels

To illustrate N-of-1 trial design features, let us consider a hypothetical study aiming to investigate the effect of physical exercise regimens for improving glycemic control. As a rationale, exercise training represents a modifiable target to reduce hyperglycemia and to mitigate type 2 diabetes (T2D) (Pan et al., 1997; Diabetes Prevention Program Research Group, 2002; Tuomilehto et al., 2001) by increasing insulin action in the skeletal muscles acutely for 24-48 hours (Sylow et al., 2017; Sylow et al., 2019). Because repeated acute exercise bouts are believed to initiate an array of physiological adaptations, repeated bouts of exercise over longer periods may further alleviate T2D (Sylow et al., 2017; Sylow et al., 2019). Estimates of the average effect of exercise training on insulin sensitivity from studies of individuals at risk of or diagnosed with T2D have been inconsistent (Maturana et al., 2021; Jelleyman et al., 2015). The continued increase in the incidence of T2D also suggests inadequate translation of these strategies. Previous research has speculated that these heterogeneous results may arise from differences among the individuals studied or in the underlying populations studied (Boulé et al., 2005; Solomon, 2018), suggesting that evaluating the effect of exercise programs for each individual might provide added benefit. In addition, each individual might benefit from a personalized program that could be evaluated by an N-of-1 trial.

An N-of-1 trial provides several advantages relative to a group randomized trial. First, it can determine the effectiveness of each type of exercise for each individual, and then return the results to the individual and their clinician to facilitate clinical decision-making. Such studies of clinical effectiveness primarily focus on whether the exercise intervention has a beneficial effect in daily life. By designing in more controlled settings, N-of-1 trials could also be used to study the biological mechanisms by which an intervention works through a careful selection of an appropriate intervention and outcome with mediators and covariates that determine a potential causal pathway. In addition, an N-of-1 trial can be designed individually to accommodate the preferences and needs of an individual participant. These preferences might include the types of exercise compared; the environments in which they are carried out; the frequency, intensity and duration of the exercise; the methods and timing of data collection; and the outcomes for which the exercise will be evaluated. Such a personalized protocol can help to optimize clinical decision-making for the participant and the clinician. In such protocols, individuals may evaluate one self-selected exercise protocol compared to their standard routine in one N-of-1 trial, or also sequentially evaluate a few different protocols. If, however, the aim is to identify an optimal protocol, then adaptive trials would be required (see section 3.5). In contrast, standard RCTs may evaluate treatments that are standardized to individual characteristics as well (Hetherington-Rauth et al., 2020), but are still limited by their aim to estimate population-average treatment effects. Hence, there is a fine balance of studying flexible interventions in traditional RCTs compared to individual interventions in N-of-1 trials that may not be combinable, and that can further be combined with individualized outcomes in fully individualized protocols. If individual inferences are of interest, then N-of-1 trials are required independent of other considerations discussed below. Going beyond the individual level, a traditional randomized trial would typically be used to answer the research question asking which physical exercise intervention is best for improving glycemic control on the population level. But one might also consider running a series of N-of-1 trials and combining results across the individuals studied. When might each of these designs be preferred?

Consider comparing three possible physical exercise interventions (going for a walk, intensive resistance training, or intensive interval training), and that each of these interventions can be done at different frequencies, intensities and durations. Different hypotheses can be considered, for example, investigating the qualitative impact of vigorous intermittent lifestyle physical activity (Stamatakis et al., 2022) or comparing qualitative differences between exercises. This would be based on the assumption that the health benefit of different interventions can be compared even if performing one intervention spends 250 kcals and performing the other spends 1000 kcals. In the following, we consider these interventions to be matched for energy expenditure. A researcher might be interested in the immediate same-day effect of different exercise regimens on glucose concentrations measured for example by an oral glucose tolerance test (OGTT), an oral glucose insulin sensitivity index (OGIS; Mari et al., 2001), time spent in hyperglycemia measured by continuous glucose monitoring or might want to know how they affect a marker such as hemoglobin A1C (HbA1C) that is hard to measure frequently and reflects longer-term changes after three months.

A series of N-of-1 trials where groups of participants are randomized to different intervention sequences in which they either do or do not perform the exercise might be applicable for measuring the average differences between exercise and no-exercise phases on glycemic control measured with OGTTs. A traditional randomized controlled trial (RCT) could also be used but would require a standardized protocol that could not accommodate personalization. The series of N-of-1 trials might also increase efficiency if the information from the repeated measurements on one individual provides information both about the treatment effect for that individual and also those for other individuals. This assumption of treatment effect exchangeability may not only increase the efficiency with which each individual's treatment effect is estimated but also might increase the efficiency of the overall design by reducing the required number of participants for testing the average effect at a pre-specified statistical power. In other words, it might be possible to trade off enrolling more participants by measuring each one more often.

When an outcome cannot be practically measured repeatedly or when it captures long-term changes such as HbA1C, one needs a traditional RCT that measures HbA1c twice on each individual, once before and once after the observation phase, in both the treatment and the no-treatment arms. Another and possibly more efficient design would be a crossover one in which each individual receives each intervention once in a randomized order. This design would require more time but allow intra-individual comparisons that would decrease the number of participants required (Senn, 2019).

Finally, if the aim is to assess which of the multiple treatments at which intensity and which duration is optimal, applying a standard series of N-of-1 trials or standard population-level RCTs might be practically infeasible. They would either require a long observation time to integrate crossover periods with all treatment-intensity-duration combinations (for N-of-1 trials) or would require many arms (RCTs). Instead, one might consider using an adaptive trial that modifies the treatment sequence or other design features (e.g., when to stop the trial) in response to an intermediate evaluation (Senarathne et al., 2020; Shrestha and Jain, 2021; Meier et al., 2023) of outcomes of interest during the trial. Adaptive designs can be of either N-of-1 or group formats.

These considerations show why it is important to match the type of design to the outcomes of interest. In the rest of the paper, we will focus on the types of outcomes for which N-of-1 trials are indicated.

## 3 Design Features of N-of-1 Trials

As discussed in the previous sections, N-of-1 trials are trials performed within one individual yielding individual-level treatment effects, and a series of N-of-1 trials can be aggregated for population-level inference. In the following description, we will focus on a single N-of-1 trial performed on a single individual. Later, we will discuss the planning and statistical analysis of a series of N-of-1 trials.

Fundamentally, an N-of-1 trial consists of periods during which the individual (repeatedly) crosses over between a set of treatments multiple times and during which outcomes may be measured one or more times. The treatment sequence is often separated into blocks, within which treatments are either randomized or assigned in a carefully designed systematic fashion. Figure 1 shows a particular example of a blocked design for a two-treatment trial in which treatments A and B are randomized within blocks of two periods such that each treatment is given once within each block and multiple outcome measurements are collected during each treatment period. Outcomes are usually collected at predetermined times, although collection may be missed resulting in missing observations. Given some assumptions, the treatment effect of interest may be determined by comparing the average of the outcomes under the different treatments (Piccininni et al., 2024). In the example from section 2, treatments A and B may represent going for a walk (A) and high-intensity interval training at shorter duration to match the energy expenditure of the walking intervention (B), each performed (e.g., daily) over respective periods of 7 days, leading to a total trial length of 42 days in this ABBABA design. As an outcome, glucose tolerance would be assessed every day, for example, by means of continuous glucose monitoring or oral glucose tolerance tests. As another alternative, exercise interventions might be performed only once per week and one OGTT weekly might be collected on the day following the exercise bout, which would limit the burden on participants and decrease possible carryover, but require a trial that is seven times as long to collect the same number of observations.

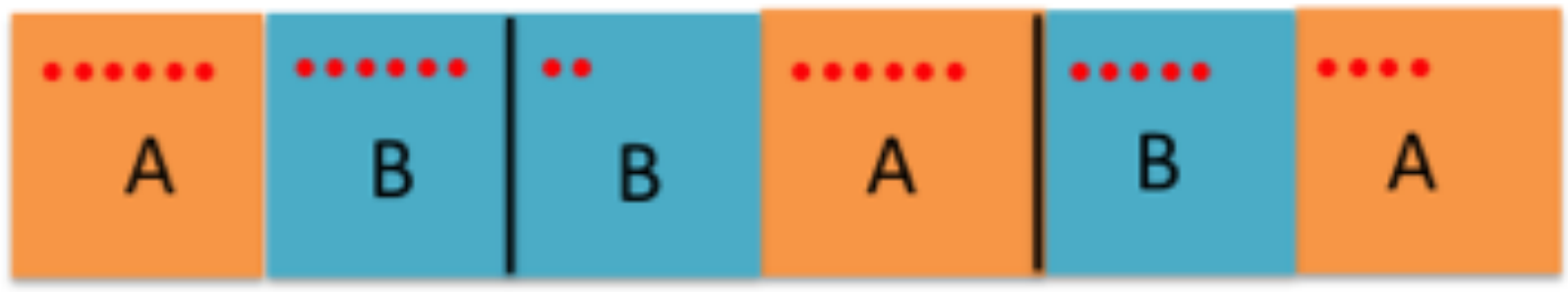

*Figure 1. Design of a two-treatment N-of-1 trial. Each colored square consists of a period during which the individual receives a specific treatment, either A (orange squares) or B (blue squares) in a randomized order. These randomized periods may be assigned within time blocks (separated by black lines). During each treatment period, outcomes may be collected on multiple occasions, although the number of measurements may vary within periods (red dots).*

***3.1 Replication and Sample Size*** A key element of trial design is determining the sample size required to achieve a statistically significant result with a high probability given an anticipated clinically meaningful treatment effect (see Section 4.9), which can be directly relevant to participants of the trial. The sample size in general refers to the total number of available data points. For a parallel-group RCT, the sample size relates to the number of participants in each arm where each participant contributes one data point. For a single N-of-1 trial, the sample size is the number of measurements obtained from that individual, which is determined by the number of treatment periods and the number of observations taken per period. Once a sufficient number of measurements on the participant across the different conditions have been made, it is possible to make inferences about the effect of each condition on that participant – hence the name *N-of-1 trial*. These measurements then help to estimate within-participant as well as between and within-period variation of the outcome variable, crucial for making proper statistical inferences. In order to increase statistical power in a trial with only 1 participant, either the number of measurements within each treatment period or the number of treatment periods must be increased. Because the nature of the trial may constrain these elements, potential sample sizes may be limited although personalized designs may permit some flexibility. In the PREEMPT trial (Barr et al., 2015; Kravitz et al., 2018), participants could choose the number of times they wanted to receive each of two treatments (2-4 times) and the amount of time they wanted to receive a treatment when it was given (1-2 weeks) so that trials could theoretically last between 4 and 16 weeks on each participant (although the total length was capped at 12 weeks per person to minimize dropout). Each participant reported several different daily outcomes so that between 28 and 84 individual outcome measurements were planned.

When a series of N-of-1 trials is planned to estimate effects in a population, the number of participants also becomes important to estimate the between-participant variance. Balancing these different components complicates sample size calculations compared to the standard parallel-arm trial in which the only consideration is the number of individuals needed for each treatment. Often, trial budgets limit the total resources that can be allocated. N-of-1 trials provide some benefits, though, because they offer choices about resource allocation. When limits on trial length constrain the number of measurements to be made on each participant, it may be possible to recruit larger numbers of participants so as to increase the total amount of information. Conversely, if the number of participants to be recruited is limited, then each trial may be extended. Careful balancing of the number of participants, the number of treatment periods and the number of measurements within treatment periods will be required to optimally allocate information in light of the expected size of the different components of variance and practical considerations of feasibility. As a rule, if responses vary more between rather than within participants, resources should be allocated to recruiting more participants but studying them for shorter periods of time; but if the within-participant variation is larger, then resources should be focused on studying fewer participants for longer times.

An R Shiny tool for determining these sample size components based on linear mixed models is available at https://jiabeiyang.shinyapps.io/SampleSizeNof1 (Yang et al., 2021). If the study aims to only investigate individual-level effects, then the tool only needs to determine the number of measurements required for each individual's treatment estimate to be sufficiently precise. This depends only on the variation expected among measurements taken on the individual (i.e., the intra-individual variation). But to investigate both individual-level and aggregated population-level effects, the sample size also involves the number of individuals to be studied and so we need to determine the number of potential randomization sequences and the number of individuals assigned to each sequence. Assuming a balanced design, namely that each treatment sequence has the same number of individuals assigned to it and that each treatment period has the same number of measurements, the tool uses a numerical search algorithm to determine the number of sequences, individuals per sequence, treatment periods and measurements per period to achieve a specific level of power and precision for estimating the population-level treatment effect. In essence, the tool either fixes the number of participants and solves for the number of measurements or fixes the number of measurements and solves for the number of participants and then determines the optimal combination from these choices. In both cases, the main requirements are estimates of the treatment effect as well as the within- and between-participant variance and the correlations among the measurements.

As an example, consider calculating the required sample size to achieve sufficient statistical power for estimating a change in OGIS in the exercise study from section 2. Assume we have pilot data showing that the intervention with the smallest effect reduced OGIS by an average of 14 ml/(min·m$^2$); the standard deviation of the individual treatment effects is 42 ml/(min·m$^2$); the standard deviation of the measurements in the control group is 46; and the the within-individual standard deviation is 23 ml/(min·m$^2$). Then, assuming a correlation of 0.5 between the intervention effect and the control means and equal correlations of 0.4 between pairs of observations on the same individual, we find that we can achieve 80% power with a type 1 error of 0.05 for different combinations of numbers of individuals and measurements. The number of measurements needed per individual decreases slightly if more individuals are studied but the difference is small because the results vary much more between, rather than within individuals. The accuracy with which each individual effect can be estimated also needs to be considered, however, and this increases with more measurements per individual. Details of using the tool can be found in the Appendix.

***3.2 Blinding*** If the study aims to determine the causal effects of a medical or biological intervention, it is important that participants and others evaluating their treatment and outcomes should remain blinded to the assigned treatments if possible. Participant blinding is particularly important in N-of-1 trials because participants are often involved in both the design and collection of the data and the personalized nature of the study may introduce a strong motivation to manage its conduct to optimize outcomes. Curiously, because individual treatment preferences are primary, the format of blinding may require more care than usual. For example, a common way to blind participants to the treatment received in a drug trial is to encapsulate the treatment in different color pills. Say a particular individual prefers blue color pills to red color pills and so has better outcomes when taking the blue pills. This preference is then real for this participant and may be incorporated into future treatments for that individual so that the effect may be sustained. However, it would not be generalizable to a population of participants and so should not be part of a generalized recommendation. Thus, care should be taken to avoid methods of blinding that may be related to treatment preferences. While blinding provides many benefits in drawing statistical inferences after the trial, the nature of the intervention often precludes it for practical reasons in N-of-1 trials. In the example of using exercise to control glucose from section 2 for instance, the nature of the interventions makes it impossible to blind participants to the intervention and may also unblind study coordinators. Without blinding, it is still possible to draw unbiased and unconfounded inferences about the effect of the intervention, however, the nature of the intervention changes. In particular, the intervention now includes the participant's knowledge about what intervention is being given and so the total effect also includes the effect of that knowledge. As an example, a participant might respond better to high intensity interval training than to walking partly because they were more motivated to do something different. In essence, the intervention now has two components and the psychological one might introduce heterogeneity into the measured treatment effects.

***3.3 Carryover and Washout*** An important consideration in crossover designs is treatment carryover when the effect of a previous treatment endures into the next period after a new treatment starts. The treatment effect in the new period then combines the effects of the treatment from the previous period with the new treatment in the current period. This is particularly the case for longer-acting treatments; studies of acute regulation are less affected by carryover. While recently, more complex causal models that incorporate carryover have been advanced (Piccininni et al., 2024; Liao et al., 2023; Gärtner et al., 2023), it is simpler to design the study to eliminate carryover and avoid the necessary modeling assumptions if possible. Such designs must of course provide proper theoretical or empirical support to buttress the assumption of no carryover.

The standard design to remove carryover builds in a washout period during the crossover between treatments. The washout can be a period during which no treatment is given, and the effect of the first treatment is allowed to disappear before the next treatment is started. The appropriate length of the washout is determined based on scientific considerations such as the half-life of a chemical treatment. Withdrawing all treatment during washout, though, can harm participants and so is often both unethical and counterproductive to inducing participation in the study. A compromise introduces an analytic washout period during which outcomes are either completely or partially disregarded in the analysis. This may be accomplished by modeling the washout trajectory (Gärtner et al., 2023). One method downweights outcomes during the washout period, ideally in a manner reflecting biological considerations. For instance, outcomes taken on the first day of a one-week treatment could be disregarded for analysis. Such an analytic washout period can be particularly useful when treatments take time to both wash out of and take effect because both effects can then occur in parallel, rather than having to be processed in series one after the other.

In our example with daily interventions and OGTT measurements, carryover would occur if the exercise regimen had an effect on glucose control that lasted more than one day. In such a case, introducing a washout period between each intervention would circumvent carryover, but would extend trial length. Alternatives would be to assume a specific model for the carryover in the analysis, or to change the study design and apply interventions and subsequent OGTT only once per week. For all considerations, it is very helpful to have insights into the length and form of carryover from the literature or pilot studies.

***3.4 Randomization and Counterbalance*** For the following, consider again the exercise study design comparing the effect of going for a walk (A) with intensive interval training matched for total energy expenditure (B), now with a design including four 7-day periods. Each period may be assigned to treatment A or treatment B. This leads to 16 possible intervention sequences that individuals might follow in their trial. We can divide these into three sets based on the maximum number of times one treatment period appears in the sequence:

1. AAAA or BBBB
2. BAAA, ABAA, AABA, AAAB or ABBB, BABB, BBAB, BBBA
3. ABAB, BABA (alternating), ABBA, BAAB (counterbalanced), AABB, BBAA

Some of these sequences would be less informative than others in estimating the difference between A and B. For instance, the first set only examines one of the treatments and so cannot inform the comparison. Therefore, randomly choosing one among the complete set of the sequences makes little sense. Instead, one might choose either to randomize individuals to one of a restricted subset of these sequences or might choose to use one or more sequences in a predetermined fixed way, often for practical or personalized reasons. The major purpose of randomization is generally to balance treatment assignment with respect to external factors that might be associated with outcomes. It also provides a sampling mechanism that justifies the application of standard statistical inference procedures (Cox, 1958).

Treatment periods can be randomized according to several different schemes characterized by the subset of possible sequences involved. As noted above, a sequence containing only one treatment cannot estimate a treatment difference, so potential randomization sequences typically disregard the first subset of all A or all B. The most common designs ensure that each treatment is used the same number of times so subset 3 is of most interest. This subset can be divided into three groups. The group {AABB, BBAA} is less useful because each sequence includes only one crossover, and the treatment difference is confounded with time. The group of alternating treatment sequences {ABAB, BABA} can be described as randomizing the first treatment, following with the other treatment and then repeating the same order. This design may be useful if it is desired to alternate treatments. The alternating treatment design was used for a study comparing two diets as a treatment for inflammatory bowel disease (IBD) among children (Kaplan et al., 2022). Four treatment periods of 8 weeks (two on each diet) were planned. One of the diets (treatment A) was much stricter, and parents were concerned that their children might not be able to adhere to the strict diet for 16 consecutive weeks as would be required for children receiving the sequence BAAB. Analogously, in our previous example, walking (treatment A) might be much easier for most people to adhere to than interval training (treatment B) which requires more motivation and equipment. Other examples in experimental physiology might be when occlusion training is compared to regular exercise, or any other pair of exercises where one is considerably more difficult and strenuous. As a result, in such situations including the study by Kaplan et al., treatments can be assigned in blocks of two periods with each participant receiving treatments in the same order in each block to prevent the same treatment from occurring consecutively. That is, all participants were randomized to alternating designs of ABAB or BABA. This principle of randomizing participants to either sequence of ABAB or BABA was also employed in a study by Vetter et al. (2024), where participants compared different anti-stress interventions in their effect of reducing stress. The third group {ABBA, BAAB} is sometimes called a systematic counterbalanced design. This design avoids the problem in the alternating treatment design that treatments always follow each other in the same order so that treatment effects may still be confounded with time. Fewer issues arise when more than four treatment periods are possible because the number of randomization sequences is increased and therefore are less likely to be confounded with time or other external features that might be unrecognized.

But valid inference on both individual-level and population-level treatment effects is still possible under some assumptions even when using fixed predefined treatment assignments that do not randomize the intervention sequence within or across participants (see Piccininni et al. (2024) for a detailed discussion). Several scenarios might argue for the use of fixed sequences. For example, if participants need to be first observed while following the standard routine A and more than one crossover is not feasible, the only appropriate design would be a fixed sequence of AB for every participant. Under such a design, and given appropriate assumptions, valid inference can still be performed when sufficient "comparable" repeated observations are available under both conditions. In exercise science, responses to both acute and chronic training often appear non-linear and time-varying, for which a first requirement is to use valid biomarkers and provide a strong rationale why the biomarker is of interest. For instance, in a one-year training program, the most significant improvements in aerobic fitness typically occur within the first three months before plateauing (Arbab-Zadeh et al., 2014). Similar patterns are observed in resistance training (Steele et al., 2023). An example from acute exercise is the reduction in blood pressure, which tends to be most pronounced within 20-30 minutes after exercise cessation but can still be observed 12 to 24 hours post-exercise, known as post-exercise hypotension (de Brito et al., 2019). In these situations, and more generally, we recommend drawing an underlying directed acyclic graph (Hernán and Robins, 2020) to map causal assumptions on the treatment effects first and then define the intervention effect of interest – for example the immediate exercise effect onto blood pressure after 20-30 minutes. Then, causal inference can be performed using more complex methods, generally requiring a higher number of data points (Piccininni et al., 2024).

Another common design feature is block randomization, in which periods are divided into blocks of an equal number of periods within which each treatment is assigned an equal number of times. The most common block type for a two-treatment study is a block of two periods in which A and B are each assigned once. The set of sequences {ABAB, BABA, ABBA or BAAB} are all block randomized. Blocking keeps the number of treatment sequences balanced throughout the trial.

Trials with more periods might employ other considerations for defining the set of allowable sequences. Kravitz et al. (2020) designed a study in which participants were assigned to a selected activity or a control three times each day over six consecutive five-day periods. Out of the 20 possible sequences, they decided to use only the 12 that included at least three crossovers and did not begin with two consecutive control periods in order to minimize dropout. Avoiding consecutive usual routine sequences at the start of the trial, though, led to more control treatments appearing in the latter part of the trial which had other consequences for analysis.

***3.5 Adaptive Designs*** In the exercise example, interest may focus on identifying the best of three or more treatments. This could be done in a sequence of N-of-1 trials, where first treatment A and B are compared followed by a comparison of the better treatment of A and B to C. For example, walking could first be compared to intensive interval training. Then, if intensive interval training is better at improving glucose control, it might be next compared to intensive resistance training. This sequence of trials, however, would take twice the time of a single trial and still might not answer the question how intensive interval training compares to going for a walk.

As a solution, recent literature has proposed methods for adaptive N-of-1 trials (Shrestha et al., 2021; Senarathne et al., 2021, Meier et al., 2023). Here, optimal designs can be derived by evaluating and comparing treatments at prespecified decision times during the trial, and then assigning the next treatment sequence taking this intermediate evaluation into account. Different strategies have been proposed for learning such optimal treatment sequences. One line of approaches are inspired by methods from reinforcement learning such as Thompson sampling. In these approaches, interim analyses are performed, for example by updating estimates of the posterior distribution of the treatment effect parameters. These can inform the probability that each of the possible interventions is best, which can in turn be used as allocation probabilities of each intervention in the next study period (Shrestha et al., 2021). Interventions that are less effective will thereby be selected less often, while interventions that are more promising will be selected more often. This "exploitation" of already obtained insights is balanced with an ongoing "exploration" and probing of all possible interventions. This carefully designed balance of exploring all possible interventions and exploiting the obtained knowledge to obtain more observations for promising interventions facilitates acquiring more informative observations for a possibly large number of exercise interventions (Meier et al., 2023). Post-study inference can then estimate and compare the effect of each intervention, thereby allowing insights into optimal interventions.

# 4 Statistical Analysis of N-of-1 Trials

N-of-1 trial data are structured as time series with measurements taken over consecutive time intervals divided into periods during which two or more interventions are applied in some predetermined sequence. As the primary purpose of the trial is to compare responses under the different interventions, the analysis of the data focuses on the treatment comparison accounting for features of the data arising from the study design that might potentially confound or bias the comparison. These features include trends that might occur over time, autocorrelation among longitudinal responses, carryover, blocked randomization, heterogeneity of treatment effect and any confounding effects present in the absence or failure of randomization.

After highlighting the importance of defining estimands of interests and embedding any analysis into a causal framework, we discuss modelling data from a single individual and then extend these models to data from meta-analysis of data from multiple individuals. Before discussing models, it should be pointed out that many single-participant data series do not use formal statistical procedures but opt instead for non-statistical graphical approaches to discern patterns. While such approaches can be useful when data series are short or when the number of crossovers is small, they become less informative and more prone to misinterpretation as the amount and complexity of the data increase. Statistical models help to quantify the size of treatment effects and to measure their precision so that appropriate statistical inferences and predictions can be made.

All of the models presented here can be fit as generalized linear mixed models either in a frequentist likelihood-based framework or in a Bayesian one. We do not discuss the technical details of these approaches as these are discussed extensively elsewhere in the literature (Pinheiro and Bates, 2000; Gelman et al., 2013). We also do not aim to provide an exhaustive overview of all proposed methodology for modeling N-of-1 trials and single time series and that there exists further work on modeling strategies using time series modeling, nonparametric approaches, or others. However, we do introduce the Bayesian approach later since we prefer it for reasons given below and it may be unfamiliar to some readers.

Interested readers may also consult other papers that discuss modeling individual treatment effects. With a focus on nutrition, Lolli et al. (2025) place N-of-1 trials in the context of cross-over trials with a particular aim to understand different types of within and between-individual and treatment response heterogeneity. Swinton et al. (2018) provide a review of statistical concepts for interpreting individual response to interventions focusing on personalized exercise and nutrition. Senn (2024) provides an informative tutorial for the statistical analysis of continuous outcomes arising from N-of-1 trials designed as paired cycles.

***4.1 Estimands and Causal Inference*** Before applying any of the statistical models described below, it is crucial to first define the estimands of interest, which are the functions of the outcomes that define the intervention effects. This is true for any trial and can be a general recommendation for the fiel of Experimental Physiology. Estimands might focus on conditional or marginal effects, on individual or average treatment effects, and may either be time invariant or may vary over time. As two examples, the conditional average treatment effect between periods of treatment and no treatment in one person might be of interest, or the mean difference in glucose levels between two treatment groups (Piccininni et al., 2024). Along with a definition of estimands of interest, a causal inference framework needs to be established that, for example, defines the superpopulation for which generalization is of interest. After defining the relevant estimands of interest, estimators such as sample mean differences, can be derived for their estimation based on assumptions about the treatment effects that should be explicitly stated. Recent work by Daza (2018), Daza et al. (2022) and Piccininni et al. (2024) discuss the causal interpretations and estimators of these estimands. In general, the theory of causal inference based on these frameworks is still in an early stage of development, and its connection to established statistical models such as those described below remains incomplete.

***4.2 Treatment*** We can set up a statistical model for calculating these probabilities that incorporates the basic design illustrated in Figure 1. Assume data generated from a design comparing L treatments (denoted A and B in Figure 1) using K blocks (separated by the black lines in Figure 1, i.e., K=3) of T=2 periods each with M measurements per period (illustrated by the red dots in Figure 1). Let $Y_{kpm}$ be the outcome for measurement *m* taken in period *p* of block *k* when treatment $X_{kpm}$ is applied. Since the treatment is the same for any two measurements *m* and *m'* in the same period, $X_{kpm}$ = $X_{kpm'}$. To simplify notation, assume two treatments so that L=2 and let $X_{kpm} = 0$ for the control or reference treatment and 1 for the treatment of interest. A valid statistical model needs to incorporate several different components: treatment effects, time trends, correlation among sequential measurements, carryover and effects from external factors. To simplify notation initially, assume no block or period effects other than the treatment given so that the outcome and treatment variables can be written as $Y_t$ and $X_t$, respectively, where t indicates the time of measurement.

We begin with the simplest model with no time trend and independent measurements

$$Y_t = \alpha + \delta X_t + \epsilon_t$$

where $\alpha$ is the mean response for the reference treatment periods (i.e., mean glucose levels under the control condition), $\delta$ is the treatment effect or difference in mean response between the two treatments (e.g., mean difference in glucose levels between the exercise condition and control condition) and the errors of measurement $\epsilon_t$ are assumed as usual in a linear regression model to be independent and identically normally distributed with mean 0 and variance $\sigma^2$. This simple model reduces to a t-test comparing the mean responses during the periods receiving each treatment. It has been used quite frequently in N-of-1 applications (Gabler et al., 2011).

***4.3 Autocorrelation*** More realistically, one might assume that the sequential measurements are correlated with each other, particularly if they are taken soon after one another. Such correlations would violate the assumption of independent measurements in the model above. The simplest type of correlation structure is one that assumes consecutive measurements share a correlation $\rho$ which declines exponentially over time, a so-called first-order autoregressive (AR1) structure. This modifies the equation above so that the correlation between $\epsilon_t$ and $\epsilon_{t'}$ at times $t$ and $t'$ is $\rho^{t-t'}$ with a magnitude determined by the time interval between the two measurements. When the measurements are consecutive, the correlation is $\rho$. A mathematical representation of this AR1 model has

$$\epsilon_t = \rho\epsilon_{t-1} + u_t$$

where $u_t$ is a normally distributed residual error after accounting for the autocorrelation.

Often, though, outcomes may be correlated in a more complex way than the simple AR1 structure can represent. N-of-1 data are in essence time series and so can be modelled with time series models These include higher-order autoregressive structures in which residuals depend on those at higher-order lags, moving averages, distributed lag models with lagged covariates and dynamic models with lagged outcomes (Greene, 2018; Brockwell and Davis, 2016; Schmid, 2001). While such models have not been typically discussed in the N-of-1 literature, Liao et al. (2023) present N-of-1 models with $k^{th}$ order autoregressive structures for residuals and with a distributed lag structure in which the current value of the outcome depends on both the current exposure and previous exposures.

***4.4 Trend*** In addition to treatment crossovers, outcomes may change because internal or external forces are changing with time. Although explicitly incorporating these forces into models by considering causal effects would be ideal, they are often unknown. Instead, one can model time trends by adding linear or nonlinear trend to the regression model. For instance, a linear time trend could be modeled by adding the term $\gamma T_t$ where $T_t$ represents the day in the study at which measurement $Y_t$ was taken. One could also add in other regression terms to model other factors that might be changing over the course of the trial.

Another way to capture changes over time assuming treatments are randomized in blocks is by adding a term for block to the model to capture changes in outcomes that occur between blocks. Such changes could simply reflect the passage of time in which case a block term represents a categorical time variable. However, an outcome can change across blocks for reasons not completely related to the passage of time but perhaps reflecting the effects of repeated treatment crossovers. This may introduce discontinuity into the evolution of an outcome. As a result, as noted earlier, variability across repeated outcome measurements taken within a treatment period may differ substantially from variability measured from repeated administration of different treatments in multiple crossovers. Different models of variability over time may be constructed: using continuous time, using blocks representing a categorical (stepped) time effect, or using a continuous time effect within blocks, a type of block*time interaction. Adding an interaction of treatment with these forms of time also permits determining whether the effect of treatment is evolving over time, perhaps increasing from learning or decreasing from habituation. We agree with Lolli et al. (2025) that further research, particularly empirical research, is needed to resolve this issue.

***4.5 Carryover*** The final feature of N-of-1 data that may need to be incorporated into models is carryover. As discussed earlier, it is simpler to design a study to avoid carryover and thus having to model it. But sometimes this may be impossible, particularly when the carryover may persist well into treatment periods. In such cases, the carryover can be modeled by comparing responses at the times of crossover and introducing model terms that describe differences from the average treatment effect that might be associated with transitions from treatments A to B or B to A. When the number of crossovers is small, carryover will be hard to model for a single individual, although with meta-analysis it may be possible to estimate population level carryover effects.

Liao et al. (2023) suggest incorporating lagged treatment terms $X_{t-k}$ into the model to estimate the effect of treatment k time units in the past on the current response. In this approach, the effect of carryover is captured by the sum of all the lagged treatment terms. In a treatment period, the carryover effect is added to the effect of the treatment being given on that day; in a non-treatment period, the carryover effect describes the total effect of treatment. In essence, this approach approximates the functional form of the decaying treatment effect as a step function. More exact representations may be used if the scientific process is better known, as for example if the pharmacokinetic properties of a drug may available.

Another approach sometimes taken is to discard the initial observations taken after the crossover to a new treatment during the time interval when the carryover might be in effect. In essence, the times discarded form an analytic washout period (Schmid and Duan, 2014). When outcomes are autocorrelated, it is important to maintain the time structure of the deleted observations in the autocorrelation structure just as if they had been missing. For instance, if the first measurement in each treatment period is discarded to allow for one day of potential carryover, the correlation between next (second period) measurement and previous measurement (last in the previous treatment period) is a lag-2, rather than a lag-1, correlation. Further models that have been applied to N-of-1 trial time series data include dynamic modeling (Vieira et al., 2017).

The potential for carryover provides another reason to increase the number of crossovers and not rely solely on within-period replication to measure treatment effect variation. The variance of a carryover effect can only be estimated through multiple crossovers and the precision with which this variability is estimated increases with the number of crossovers. Unlike in a meta-analysis where crossovers may accumulate across participants, in a single N-of-1 trial, the number of crossovers can only be increased by adding them on. Without multiple crossovers, it is impossible to distinguish variability caused by replication heterogeneity from that caused by re-introducing a treatment that had been previously withdrawn.

***4.6 Meta-Analysis*** We noted earlier that we could estimate average effects in the population and in subgroups and potentially get better estimates for each individual by combining trials from different individuals. This can be achieved in a meta-analysis, which has been largely applied in the context of summarizing the evidence of a given research question based on a systematic review of the relevant literature of, for example, standard population-level RCTs. Meta-analysis can be employed similarly in the context of N-of-1 trials where, instead of aggregating the effects observed in different RCTs, multiple N-of-1 trials are aggregated.

We can extend the model for a single individual trial to one for multiple trials from different individuals, first adding notation to label each individual trial. Using a model that incorporates a linear time trend and autocorrelation, we add a subscript *i* to label individuals so that

$$
\begin{aligned}
&Y_{it} = \alpha_i + \delta_i X_{it} + \gamma_i T_{it} + \epsilon_{it} \\
&\epsilon_{it} = \rho_i \epsilon_{it-1} + u_{it} \\
&u_{it} \sim N\left(0, \sigma_i^2\right)
\end{aligned}
$$

describes the model for individual *i*. Note that individuals have their own unique parameters $\alpha_i$, $\delta_i$, $\gamma_i$, $\rho_i$ and $\sigma_i^2$ to describe their own effects. To learn across individuals, we must relate these individual parameters to each other.

Three different types of relationships among the parameters have been suggested (Schmid and Yang, 2022). One assumes that individual parameters are separate and unrelated; the second assumes that individuals share common parameters; the third assumes that individuals share parameters from a common probability distribution. Each has advantages and disadvantages.

Unless parameters are related in some way, it is impossible to learn anything about one individual from others. Thus, it would seem that the first approach is no better than modeling each individual separately. However, it might be useful to keep some types of parameters unrelated but relate others. For example, the model intercepts $\alpha_i$ describe individual responses when other model factors are not in operation (or are set to their reference levels). If the population of individuals is very heterogeneous or is not chosen in a random manner, these intercepts may be quite variable and specific to the circumstances by which individuals entered the study and their individual characteristics. In such cases, it might be better to let the intercepts be unrelated and estimate each separately. These are often called fixed effects.

Sharing a common parameter is a strong assumption. For example, a common treatment effect $\delta_i = \delta$, implies that treatment has the same effect on all participants. This might be approximately true if the participants are very similar but is unlikely in a heterogeneous group. Assuming that the correlation $\rho_I$ between successive measurements is similar across individuals, though, may not be an unrealistic assumption. Moreover, when data are sparse, some parameters may be difficult to estimate on each individual and an assumption of commonality might be practical.

The assumption of a common probability distribution is the most common approach to dealing with shared parameters in meta-analysis. The method assumes that each parameter is randomly drawn from a distribution of possible parameters and so the method is usually called random effects. For shared treatment effects, one would assume that each individual treatment effect $\delta_i$ comes from a common distribution, often assumed to be a normal distribution with mean $\mu_\delta$ and variance $\sigma_\delta^2$, written $N(\mu_\delta, \sigma_\delta^2)$. This is an efficient approach since only two parameters need to be estimated to describe the population distribution and the posterior distribution of the individual $\delta_i$ can be expressed in terms of these parameters and the data (Schmid and Brown, 2000).

A (partially) random effects model in which the treatment and trend effects are random effects, the intercepts are fixed effects and the correlations and residual variances are common effects can be expressed as a multilevel form as follows

$$
\begin{aligned}
&Y_{it} = \alpha_i + \delta_i X_{it} + \gamma_i T_{it} + \epsilon_{it} \\
&\delta_i \sim N(\mu_\delta, \sigma_\delta^2) \\
&\gamma_i \sim N(\mu_\gamma, \sigma_\gamma^2) \\
&\epsilon_{it} = \rho\epsilon_{it-1} + u_{it} \\
&u_{it} \sim N(0, \sigma^2).
\end{aligned}
$$

Here, $\delta$ is the average treatment effect and $\gamma$ is the average linear trend across individuals. The variation in the treatment effects and linear trends are expressed by the variance components $\sigma_\delta^2$ and $\sigma_\gamma^2$, respectively.

If more is known about some of the random parameters or a common mean is felt to be too strong an assumption, the means can be modeled themselves. For instance, if one thought that the treatment effect of males differed from that of females, a model with two means could be constructed by expressing $\delta = \delta_1 + \delta_2 Z_i$ where $Z_i = 1$ if individual i is male and 0 if female. In this case, the mean for females is $\delta_1$ and the mean for males is $\delta_1 + \delta_2$. Thus, if enough participant characteristics are known, it may be possible through such regression formulations to categorize individuals into subgroups that are homogeneous enough that a common distribution within the subgroup is a reasonable assumption while accommodating heterogeneity through the different subgroups. This approach often provides an effective compromise among the three approaches to parameter definition.

In addition to using regression to explain heterogeneity of treatment effects among individuals, various extensions to the assumption of a normal random effects distribution have been proposed to describe the residual variance of the individual study effects. These include longer tailed distributions like Student's t, skewed distributions, mixture distributions and Dirichlet processes. Although they have not been used frequently in meta-analysis to date, recent work suggests that the standard normal assumption works well unless substantial heterogeneity or substantial non-normality expressed as skewness, outliers or clusters of effects is present (Panagiotopoulou et al., 2025). In such cases, distributions tailored to the type of type of heterogeneity will work better. If such heterogeneity is expected, trying different random effects formulations to determine whether some describe the data better is recommended.

Although the random effects model is widely used in meta-analysis to generalize inferences beyond the particular studies included in the meta-analysis (Higgins et al., 2009; Borenstein et al, 2010), investigations have found that it can perform poorly when the number of studies, N, is small because the between-study variance is poorly estimated. This results in poor predictive performance in new studies as well as high type 1 error rates, low power and confidence intervals that do not have proper coverage of the true parameter resulting in incorrect inferences (Guolo and Varin, 2017; Jackson and Turner, 2017). Thus, extreme care must be used with small numbers of studies, say $N < 10$ (and random effects models should probably be avoided when $N < 5$). Generalized inferences should be made with caution unless external information about the between-study variance is known from other sources. One approach is to use a Bayesian model, which can incorporate external information to inform estimation of the variance parameter, but this approach must also be used with care unless the external information is sufficient to overcome the lack of data (for further discussion, see Section 4.8 below). It is important to qualify these findings, though, because they apply to standard meta-analysis which combine the results of clinical trials in which the units of analysis are individuals, rather than to N-of-1 trials that combine measurements in a single individual. Within- and between-study variability may be quite different in such cases. The N-of-1 trial measurements are also likely correlated, whereas measurements on different individuals in a standard trial are independent. Thus, it will be important to evaluate such claims for N-of-1 data.

Finally, it is worth noting that carryover is more easily modeled when meta-analysis allows combining the results of multiple trials. As was noted earlier, estimating carryover can be difficult in single participant trials when the number of crossovers is few. Data from a meta-analysis of trials include many more crossovers and so increase the information with which to estimate carryover effects. For example, a simple carryover model would have separate parameters for the effect of crossing over from A to B and from B to A. When these crossovers are repeated multiple times in multiple trials, formulating the carryover parameters as random effects allows estimating both an average carryover and individual carryovers which are informed by borrowing across individuals. Other more complex carryover patterns such as when carryover persists or decays over time can also be accommodated.

Combining trial results through meta-analysis provides one other key benefit. If the individual-level parameters can be related to each other across a common random effects distribution, then their posterior estimates are a weighted average of the information coming from their own measurements and the information coming from others (Schmid and Brown, 2000). If we consider the information coming from others as the average in the population, then the best (posterior) estimate for each individual's effect learns from the population to modify the information that the individual alone provides. Intuitively, if the individual provides substantial information, then we do not need to use information from the population to learn about that person; but if the information from the individual is scant (for example because few measurements were taken or many were missed), then we will learn something by borrowing from others who are alike. This is called borrowing strength from related individuals.

***4.7 Missing Data and Non-Adherence*** All trials can suffer from non-compliance or lack of adherence of participants to the study protocol. This can involve failure to perform the indicated treatment and failure to record outcomes. Non-adherence reduces the amount and quality of the data collected and can lead to bias and loss of precision of quantities of interest. Its creation of missing data values complicates data analysis. Non-compliance with study protocols or frequent non-reporting of outcomes can be especially problematic because individuals or situations for which it occurs often differ substantially from those when it does not. Ignoring the missing values by excluding them from analysis necessarily then fails to account for these changed conditions and invalidates any generalized inference. The reader can and should consult the large literature on the effects of missing data and non-compliance (Little and Rubin, 1983)

The special nature of the design, data collection and interpretation of N-of-1 trials pose additional considerations in discussing the perils of non-adherence. First, because inferences are made for individuals, non-adherence may impact drawing inference for some individuals more than for others. While missing values in a population-based study degrade the conclusions that can be drawn for the average participant, missing values in an N-of-1 trial degrade inferences for that individual. Even when data from multiple individuals are combined in meta-analysis, inferences on non-adherent participants will be most affected. Second, because in many cases individuals will be collecting and recording their own data, non-adherence to a protocol may also affect collection of all types of study data. Third, since designs are individualized, non-adherence to a specific design may invalidate any inferences for that design.

On the other hand, several factors work to reduce the effects of non-adherence. Since individuals design and help run their own trials, they will probably be more motivated to stick with the protocol, thus reducing the chance that they will become non-adherent. Also, since individuals conceptually provide enough evidence to draw inferences from their own trials, it may be possible to quantify the effect of non-adherence. The diet study referenced earlier (Kaplan et al., 2022) and discussed in Section 4.9 more fully below provides a useful example. In that study, 24 of 54 participants completed the full four period protocol, 9 completed at least two periods, and 21 dropped out before fully completing each treatment phase once. Not unexpectedly, those who completed the trial had better results than those who did not. The authors were able to conclude that those who could stick with the diet had a good chance of benefiting from it; those who could not received no benefit (and often dropped out because they were finding no benefit).

***4.8 Bayesian Models*** The models outlined so far only specify a likelihood and thus can be fit in either a frequentist or Bayesian framework. The frequentist approach typically optimizes the likelihood from the model producing maximum likelihood estimates along with standard errors (Van Dongen, 2006). Statistical packages can produce these estimates for a variety of models. The Bayesian approach combines the likelihood with external information about the model parameters expressed through their prior distribution using Bayes rule to produce a posterior distribution that describes knowledge about the parameters given the proposed model and prior (Gelman et al., 2013; Schmid et al., 2020). Such computations can be done analytically in special cases, but usually require numerical methods such as Markov chain Monte Carlo (Brooks et al., 2011)

The Bayesian approach is useful for several reasons (Schmid and Yang, 2022). Most importantly, Bayesian models produce the complete joint posterior probability distribution of all model parameters after incorporating information from the data through the likelihood and from external sources through the prior distribution. The posterior distribution describes the probability that the values of any model parameter or set of parameters falls into a specific range. This allows constructing direct statements about the probability of scientific hypotheses related to the size and ranking of treatment effects including the chance that one is larger than another and the amount of the difference. Such statements then enable the personalized decisions about care that motivate the application of N-of-1 studies. For example, results can be derived that quantify the probability that a given participant has a glucose value improved by a pre-defined clinically meaningful difference when walking or when doing intensive interval training.

Furthermore, because Bayesian methods combine information from the likelihood of the observed data with external information about model parameters expressed through their prior distributions, they allow participants to incorporate personal information external to the data into their decision making in a quantifiable manner.

Bayesian inference also offers other advantages. These include avoiding the need to rely on asymptotic normal distributional assumptions implicit when reporting only a point estimate and a standard error. Such assumptions can be problematic with small sample sizes when asymptotics do not hold and sampling distributions may be non-normal. In addition, Bayesian modeling can simplify handling of missing data by combining imputation of missing values with model estimation (Gelman et al., 2013; Little and Rubin, 2019). This can be especially important with time series models that incorporate time-specific autocorrelations in which maintaining the proper time sequence is important. Simply ignoring missing observations would spoil the sequential nature of the measurements.

In Bayesian models, posterior inferences about model parameters are constructed by combining information from the model likelihood with prior information about the model parameters expressed through their prior distribution. In a single N-of-1 trial, the model parameters are $\alpha$, $\delta$, $\gamma, \rho$ and $\sigma^2$. For a meta-analysis of a set of trials, the random effects formulation models these individual parameters as coming from distributions whose parameters such as $\mu_\delta$ and $\sigma_\delta^2$, called hyperparameters, require prior specification. When substantial information about the likely values of model parameters is available before a study, either from prior investigations or biological knowledge, incorporating this information into Bayesian models through the prior distribution of the parameters will improve the inferences that can be drawn. As an example, if response to an intervention varies in a known fashion with the time of year at which it is given, then the time when the study is performed can inform the expected effect.

In general, when the data provide a large amount of information to the likelihood (e.g., through a large sample size), posterior inferences will be robust to choice of the prior, whether they be strong priors that imply rather precise knowledge about model parameters or weak priors that imply little knowledge. But when data are sparse or more generally do not provide sufficient information to the likelihood, the posterior distribution may be quite sensitive (i.e., might change substantially) with different prior distributions. This poses no problem when a strong prior is generally agreed upon because then posterior inference will be heavily influenced by prior knowledge. However, when the data are weak and prior information is also weak, then sensitivity to choice of prior causes problems.

In general, when little is known *a priori* about model parameters, analysts often choose to use a weak or non-informative prior distribution. Common choices include uniform distributions that place equal weight on all parameter values or normal distributions with large variances which tend asymptotically to uniform distributions. Because certain forms of non-informative prior distributions lead to posterior inferences that are numerically equivalent to the likelihood, it is sometimes possible to turn a likelihood-based inference into a Bayesian formulation. However, with many common models problems arise because a prior that is non-informative for one parameter can be quite informative for a function of that parameter that may also be of interest and so it becomes impossible to ensure that all quantities of interest are robust to prior choice. Van Dongen (2006) demonstrates an example of this phenomenon, showing that placing non-informative priors on the intercept and slope of a logistic regression model results in an informative prior on the conditional probability of the expected outcome for a specific covariate value. In many problems including meta-analysis, prior sensitivity is most problematic for variance parameters, partly because their posterior distributions are bounded at zero and are skewed and partly because sample sizes are often too small for the likelihood to have information about them (Gelman, 2006).

Since using non-informative priors for variance and correlation parameters and in nonlinear models leads to posterior distributions that are highly sensitive to the specific forms of the priors, it is advisable to avoid them. Instead, one should use available information to develop weakly informative priors that place sufficient probability on all reasonable parameter values and reduce the posterior probability of improbable values for them and transformations or functions of them that are of interest. Some recommended choices include bounded uniform distributions and half-Cauchy, half-normal and half-t distributions (Röver et al., 2021; van Dongen, 2006; Gelman, 2006).

It is also important to try different prior formulations to determine whether posterior inferences are sensitive to the choice. Posteriors that are robust to choice of the prior provide confidence that inference rests on a solid foundation. Posteriors that change substantially with choice of the prior suggest that the data are weak and drawing strong conclusions is risky. One exception to this rule occurs when posterior distributions corresponding to different strong prior distributions are compared and found to lead to different conclusions. In this situation, the different posteriors represent different degrees of belief among individuals who hold strong conflicting priors when looking at the same data. Spiegelhalter et al. (1994) described such priors as enthusiastic and skeptical reflecting the potentially divergent views of government regulators and industry representatives seeking approval of a new product.

To reduce influence of the prior distribution, it is important to ensure that a trial is designed with a sufficient sample size to answer the questions of interest with high probability of sufficient precision. The sample size calculator described in Section 3 works for analyses using linear mixed models based on the likelihood. While not implicitly Bayesian, it does ensure that sufficient data are available to estimate the likelihood adequately and so should reduce sensitivity to choice of the prior in a Bayesian analysis as well.

To properly determine sample sizes for a Bayesian model, though, is more complex and requires specification of two prior distributions (Wang and Gelfand, 2002). One simulates parameter values that combined with the hypothesized model generates hypothetical responses. The other forms the prior distribution that combined with the model likelihood and the simulated responses generates draws from the posterior distribution. This approach has not to date been applied to N-of-1 trials and so is a fruitful area for future research. It can also be extended to meta-analysis (Schmid et al., 2004) and adaptive trials (Lee, 2024).

***4.9 Example*** As an example of the application and interpretation of N-of-1 trials (using Bayesian methods), Kaplan et al. (2022) described the posterior probabilities that application of two different types of diets, a specific carbohydrate diet (SCD) and a modified one (MSCD) reduced stool frequency among children with inflammatory bowel disease (IBD), either Crohn's disease or ulcerative colitis, compared to each other and to their own baseline diet in a series of N-of-1 trials. In each trial, children were followed on their usual (baseline) diet for two weeks and then were randomized to either SCD or MSCD for a period of 8 weeks. After 8 weeks, the child was switched to the other diet for 8 weeks. The diets were then repeated for two additional 8-week periods in the same order (alternating treatments) for a total duration of 32 weeks of diets. Stool frequency was tabulated daily as an integer count and was modelled as a Poisson random variable using a generalized linear mixed model with a separate intercept for each diet. Trend and autocorrelation were not included in the model. Missing values were imputed as parameters in the Bayesian models. The model was fit using MCMC; details can be found in Kaplan et al. (2022).

Here we focus on the presentation of results from the posterior distributions of the diet effects for 51 individual trials that provided stool data. Figure 2 presents the relative risk ratios (RR) and 95% credible intervals for three sets of participants. Full completers finished all 4 diet periods, early completers finished the first two but did not complete all 4; withdrawals did not finish the first 2 periods. The gray region indicates a clinically indifferent region in which the relative change was less than 20% in either direction that clinicians believed was not large enough to warrant a change in treatment. This is similar to the region of practical equivalence (ROPE) that has been suggested as a way to judge the strength of an effect (Kruschke, 2018). Such a region should ideally be determined when the study is being designed to aid in later interpretation of results as in Maturana et al. (2021). Sadly, most studies fail to define ROPE or a similar region. The top row, comparing the strict diet to the baseline, shows that the expected number of stool events is clinically significantly reduced with the strict diet (RR < 0.8) for some children who completed the trial but increased (RR > 1.2) for others. Although many of the credible intervals cross into or past the region of indifference so that the effects are generally not clinically significant, a few individuals (P10, P2, P12, P18) do show benefits and at least one (P6) appears to have more stools with SCD. The effects are less among the early completers and withdrawals. Similar results can be seen for the modified diet (second row). The bottom row shows that SCD and MSCD generally had similar effects with most estimates in the region of indifference, although P2 appears to have done better with a strict diet and P6 and P15 did worse with a strict diet. Note that we can generally not compare SCD and MSCD for many of those who withdrew early because they did not collect data on both diets.

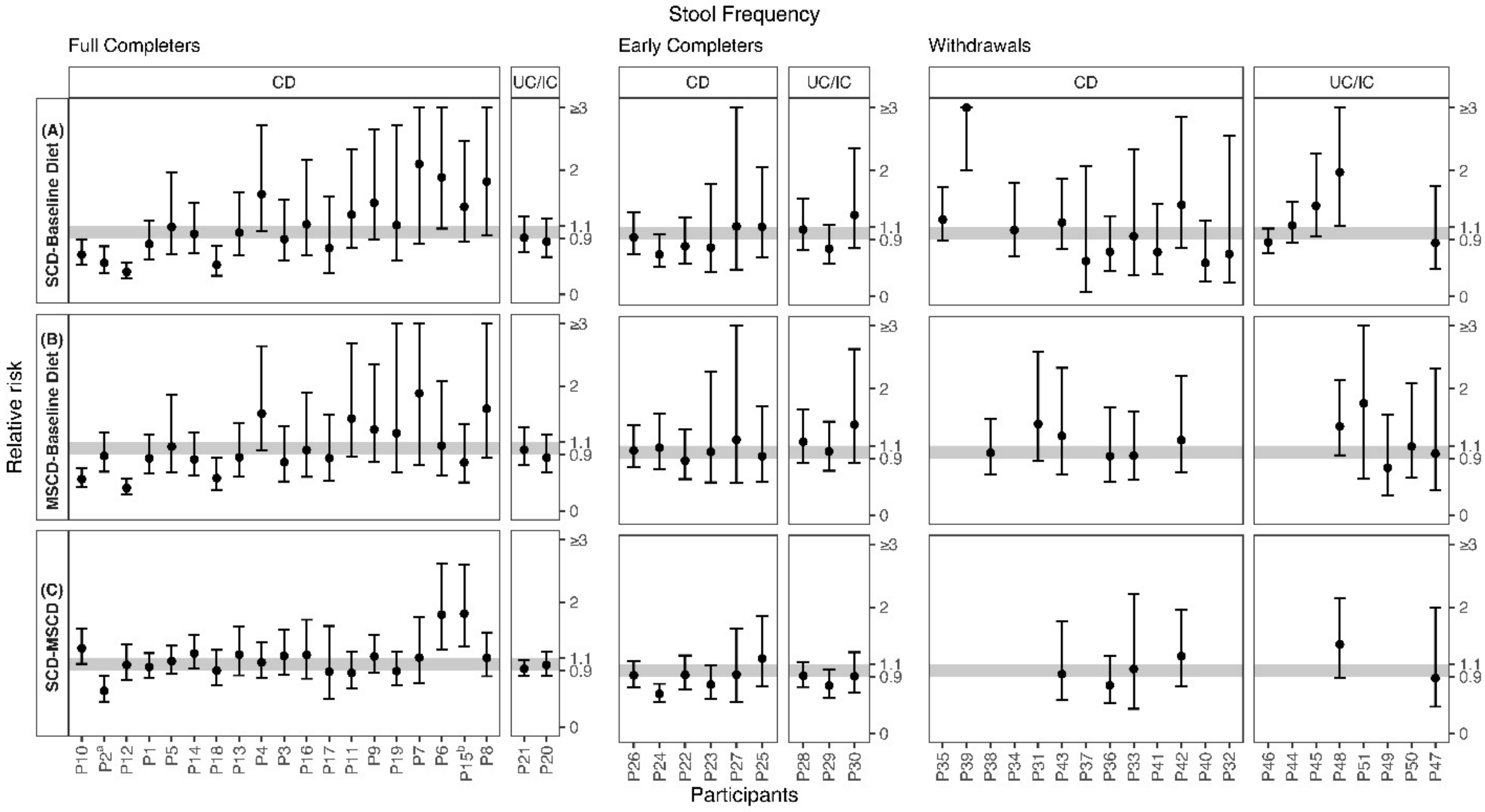

*Figure 2: Risk Ratio and 95% Credible Interval for Change in Stool Frequency in Individual N-of-1 Trials. Risk ratio and 95% Credible Interval (CrI) for full completers, early completers, and withdrawals for three diet comparisons: (A) SCD versus Baseline Diet, (B) MSCD versus Baseline Diet, and (C) SCD versus MSCD. Within each diet comparison, individual trial results are ordered by disease type and by extent of baseline symptoms (more to less). The shaded region in gray represents a difference that is not clinically meaningful (< 10% relative change in either direction). For withdrawals, participants with measurements only on baseline diet are not included in the figure. Note: a indicates a child response was used in analysis, b indicates that the participant was randomized to begin with SCD, but began with MSCD.*

Figure 3 shows results in a different way, presenting the posterior probabilities that each individual had a clinically significant benefit favoring the diet (RR < 0.8), a clinically insignificant effect (0.8 <= RR <= 1.2), or a clinically significant harm with the diet (RR > 1.2). In the top row, dark regions favor SCD and lighter regions favor baseline. The patterns noted in Figure 2 are apparent. Some individuals do significantly better on SCD, others do significantly better on baseline, and for some results are indeterminant.

The two figures complement each other nicely. Clinical significance is more easily recognizable with Figure 3, but the point estimate and region of highest probability are easier to discern in Figure 2. Similar summaries of the posterior distribution have been suggested in the literature. These include ROPE+HDI which calculates the proportion of the posterior probability in the highest posterior density interval (HDI) that falls in the region of indifference (Kruschke, 2018) and the probability of direction that calculates the proportion of the posterior distribution that has the same direction as the median effect (Makowski et al., 2019). Whenever treatment recommendations are provided based on responders/non-responders to treatment based on the posterior distribution, we recommend that cutoffs of clinically meaningful effects and cutoffs of the posterior distribution are defined and published a priori (Berg et al., 2024) in protocols, see e.g. Vetter et al. (2024) and Vetter et al. (2025). Also, it is important to carefully define what a clinically/biologically meaning effect based on prior evidence or pilot studies.

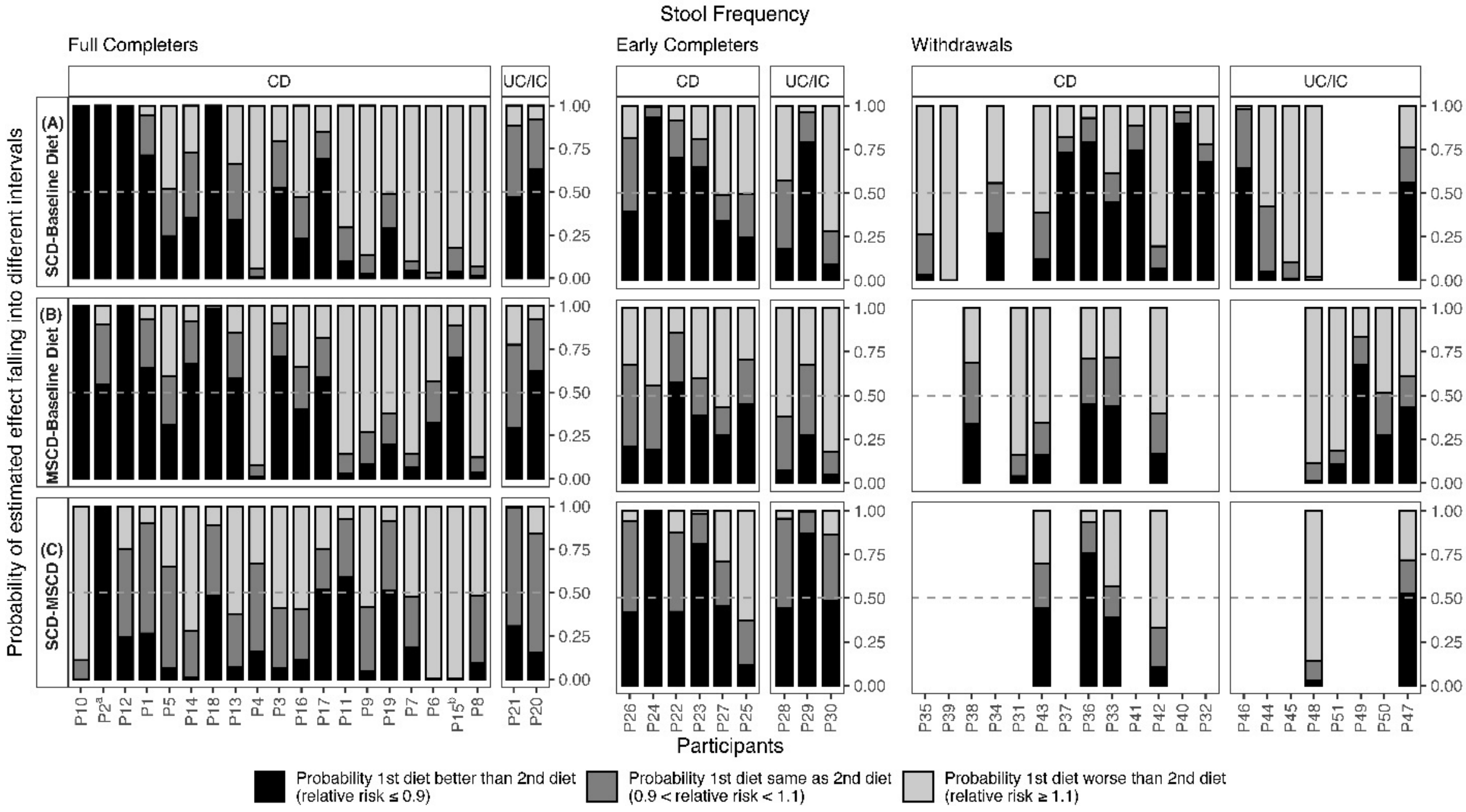

*Figure 3: Probability of Symptomatic Improvement in Stool Frequency in Individual N-of-1 Trials. Probability of symptomatic improvement for full completers, early completers, and withdrawals for three diet comparisons: (A) SCD versus Baseline Diet, (B) MSCD versus Baseline Diet, and (C) SCD versus MSCD. Within each diet comparison, individual trial probabilities are ordered by disease type and by extent of baseline symptoms (more to less). For withdrawals, participants with measurements only on baseline diet are not included in the figure. Note: a indicates a child response was used in analysis, b indicates that the participant was randomized to begin with SCD, but began with MSCD.*

## 5 Digital Features of N-of-1 Trials

Most randomized trials require extensive research infrastructure that organizes participant contact, visits and examinations at the study center at all time points, and processing and collection of samples and data. N-of-1 trials that can be run by individuals, clinicians, researchers, or combinations of them, require frequent assessments of the health outcomes of interest in order to achieve high statistical power for individual-level and population-level analyses. For investigating fine-grained questions including time-varying effects or mediators, even more data are required. This is not practically feasible within a standard trial infrastructure and has inhibited a sustainable set-up of N-of-1 trials in the past. Instead, digital apps can enable decentralized digital trials through managing the study design, data collection, data analysis and even interpretation of results. Apps can also obtain electronic consent, perform checks of eligibility, manage enrollment and randomization, monitor progress and promote increased adherence. In addition to facilitating the digital collection of study outcomes, such user-friendly tools can limit the burden on study participants who might participate in a trial over multiple weeks and months and enter data on their phone daily.

Such digital N-of-1 trials collect patient-reported outcomes by questionnaires, multiple choice questions, free text fields or performance ratings. In addition, digital tools can easily use wearables and devices to assess biometric data (e.g., blood glucose levels), include health outcomes assessed by audio or images or use technology that assesses molecular markers such as blood glucose values. As another important feature, adverse events can be tracked by patients and monitored by physicians fully remotely. Digital tools can incorporate reminders to take a given intervention (i.e., notification motivating going for a walk; notification reminding to take medication) or could provide the intervention directly (e.g., show video of yoga or physical exercise classes). Providing the interventions through a digital app also facilitates assessment of adherence to the intervention. Notifications in the form of app-related push notifications or text messages can also remind participants regularly at appropriate times to record their outcomes and adherence to the randomized treatment assignment. For example, the app might send a daily notification at 10 in the morning reminding participants to perform their physical exercise, and then send another reminder to fill out the health data at a pre-specified time after the exercise. At the end of the trial, an automated data analysis can be performed using any of the statistical methods described above and the results provided to the study participant together with a visual and textual interpretation such as the probability that the tested intervention provides a clinically meaningful improvement. As a support, instructional videos can be provided at all stages of the trial.

Different digital tools are available for recording information from study participants. Generally, tracking apps and any tool or platform that allows creation of online surveys, journaling and recording answers to questions might be used to collect information for a digital N-of-1 trial. However, while many available tools can record outcomes digitally, few tools have been developed with a specific focus on N-of-1 trials. For researchers and physicians in experimental physiology, we believe that a platform enabling and supporting the design and conduct of digital N-of-1 trials across a wide range of applications is the most beneficial.

We have developed such a user-friendly general tool for digital N-of-1 trials called StudyU that sets up a design, collects and stores data and makes it available for post-trial analysis, and interacts with users in real-time. StudyU (Konigorski et al., 2022; https://studyu.health) is an open-source and free platform that includes the StudyU Designer, a study database, and the StudyU App. With the StudyU Designer, single N-of-1 trials and a series of N-of-1 trials can be defined through a web interface by researchers or clinicians, including all relevant aspects outlined above. These study protocols as well as the recorded participant data are stored in the study database. Study participants can access these studies through the StudyU mobile App, enabling fully-digitally enabled studies. Different patient-reported outcomes can be defined. Newer developments focus on the inclusion of wearables and sensors (Daza, 2018, Daza et al., 2022, Zhou et al., 2022) and other multimedia data (Fu et al, 2023; Schneider et al., 2023), allowing passive collection of high-frequency measurements from participants. For our exercise example trial, continuous glucose monitors or other tests for glycemic control might be integrated into StudyU. After trial completion, the anonymized participant data can be extracted, published, and its results provided back to the study participant.

In addition to platforms like StudyU that facilitate digital N-of-1 trials for research collaborations between scientists and participants, researchers have developed apps that enable individuals to design and carry out N-of-1 trials by themselves without supervision. Two of these, Self-E (Daskalova et al., 2021) and StudyMe (Zenner et al., 2022), do not store participant data centrally and so cannot be used for research purposes. Instead, they focus on consumer-driven self-experimentation and self-quantification. Finally, some disease-specific apps (e.g., migraine apps; Seng et al., 2018) that record observational data could be used for N-of-1 trials, but would require manual set-up of the trial externally or a linkage to a platform like StudyU.

# 6 Discussion

N-of-1 trials, and in particular digital N-of-1 trials, provide an innovative study design that can be used to integrate individuals into research studies more directly or even put them at the center in co-designed studies. Carrying out a series of N-of-1 trials provides an alternative to parallel-arm RCTs that focuses on estimating treatment effects for individuals and not just the average person. We have laid out the main design and analytic principles behind N-of-1 trials in order to give guidance about whether N-of-1 trials can be a suitable study design for a given research question in experimental physiology and how to plan and analyze them. As an illustrative example, we have described a possible application of N-of-1 trials investigating the effect of different physical exercise interventions on glucose control and have described a method for reporting results that we used in a previous series of N-of-1 trials testing two diets for treating children with inflammatory bowel disease.

N-of-1 trials are uniquely designed to determine effects for individuals, not just groups of individuals. A properly designed N-of-1 trial can determine whether an intervention works for a particular person and how large the effect is. This may be most beneficial when treatment effects vary considerably among individuals. N-of-1 trials are most easily conducted and interpreted for conditions that are chronic and stable, for which repeated applications of the intervention can give repeated estimates of the intervention effect – but N-of-1 trials are also applicable in more complex situations (Piccinnini et al., 2024). When carried out in parallel among multiple patients, a series of N-of-1 trials can also provide evidence of the generalizability of the treatment effect through estimation of average effects across individuals and improved estimates of an individual's effect by borrowing from others like them. This is most beneficial under the assumption that pooling across participants is appropriate (Yang et al., 2021; Schmid and Yang, 2022; Araujo et al., 2016). Careful design allows for allocating resources properly between studying more individuals and studying each individual longer, thus increasing statistical power for measuring both individual and population level effects. As described in this review, there are multiple ongoing efforts in advancing methodology for N-of-1 trials and making them application to complex situations, which still need to be harmonized in future work in order to provide best practice guidelines.

N-of-1 designs are not a panacea, however. Because of the repeated crossover design, it is important that participant responses be comparable across the periods during which treatments are assigned. Therefore, a condition that resolves will not be amenable to this design. An outcome that permanently changes or carryover of effects from the previous treatment into the next treatment period also pose a challenge. The need for taking many repeated measurements to accurately estimate efficacy for an individual participant also tends to lengthen an N-of-1 trial relative to a parallel arm trial, thus increasing the risk of non-adherence and dropout. Moreover, because data collection is often carried out by participants, intermittent missing data frequently arise. These challenges might be addressed by increasing intrinsic motivation of participants since they can also benefit directly from taking part in the trials by knowing their results.

Some of the potential advantages of N-of-1 designs can also become limitations. While the capacity to design an individualized protocol can lead to better outcomes and more motivated participants, it may limit the ability to generalize findings to others. In the PREEMPT study, patients with chronic non-cancer pain were given the opportunity to compare two treatments of their choice in a personalized trial that allowed them to choose the length and number of the treatment periods (Barr et al., 2015, Kravitz et al., 2018). While the trial was popular with participants (Whitney et al., 2018) and results suggested that N-of-1 trials might potentially improve outcomes relative to standard care, the wide variety of comparisons chosen prevented any comparison of the effectiveness of specific treatments. In the IBD diet trial described in Section 4.9, many children responded to one or both diets, but many others did not. Therefore, while one could conclude that a specific carbohydrate diet or a modification of it is a useful tool for pediatricians treating IBD, it will not necessarily work for even the majority of those with IBD. This limits generalization of the findings to others, which generally is the aim of population-level research studies, but at the same time brings up the question if generalization is meaningful or if more personalized research questions should be pursued.

In addition, individuals who might participate in an N-of-1 trial might differ systematically from those who would enroll in a parallel-arm RCT. Whereas RCTs often require extensive research infrastructure that limits where trials can be conducted and who might be able to participate in them, N-of-1 trials may be carried out in community settings, thus potentially increasing their generalizability. But the lack of research infrastructure faced by many who might want to carry out N-of-1 trials limits their application. Digital implementations can increase research opportunities as outlined in Section 5, but will require care in development to reach the level of research expertise and experience enjoyed by major research centers. In addition, those motivated to design their own trials and monitor their own progress (such as members of the quantified self-movement (Feng et al., 2021)) may differ in important ways from those who might choose to enroll in a trial designed and run by a professional research team.

Thus, N-of-1 trials should be considered a supplement to, and not a replacement for, the standard parallel-arm RCT. They serve an important purpose, particularly in assessing individualized treatment, and can help to measure, assess and understand treatment effect heterogeneity. They can also serve as an important tool for individuals to participate in and take control of their own health and to better understand the scientific process.

Experimental physiology offers many potential applications where N-of-1 trials can provide value in estimating individual treatment effects or delivering results back to patients, especially with growing access to digital tools to assess physiological outcomes. Such applications include exercise interventions, pharmacological interventions, or acute environmental stressors in diseases such as hypertension, heart failure, diabetes and other endocrine or also neurological disorders with considerable heterogeneity of treatment effects. Because many physiologic processes can be monitored with sensitive instruments that provide multimodal measurements such as continuous glucose monitors, wearable sensors, and also highly detailed images, repeated measurement is feasible, thus opening up experimental physiology to efficient and powerful N-of-1 study designs.

## Author contributions

SK and CHS drafted the manuscript jointly with inputs from MRL who conceptually provided the case example with inputs from CHS and SK. All authors have read and approved the final version of this manuscript and agree to be accountable for all aspects of the work in ensuring that questions related to the accuracy or integrity of any part of the work are appropriately investigated and resolved. All persons designated as authors qualify for authorship, and all those who qualify for authorship are listed.

## Acknowledgements

MR-L was funded by The Centre for Physical Activity Research (CFAS), which is supported by a grant from TrygFonden.

## Conflicts of Interest

SK and CHS declare no conflicts of interest. MRL is employed by Novo Nordisk A/S. Novo Nordisk A/S had no role in the preparation, review, or approval of the manuscript: and decision to submit the manuscript for publication.

## References

1. Araujo A, Julious S, Senn S. Understanding Variation in Sets of N-of-1 Trials. *PLoS One*. 2016 Dec 1;11(12):e0167167.
2. Arbab-Zadeh A, Perhonen M, Howden E, Peshock RM, Zhang R, Adams-Huet B, Haykowsky MJ, Levine BD. Cardiac remodeling in response to 1 year of intensive endurance training. *Circulation*. 2014 Dec 9;130(24):2152-61.
3. Atkinson G, Williamson P, Batterham AM. Issues in the determination of 'responders' and 'non-responders' in physiological research. *Experimental Physiology*. 2019 Aug;104(8):1215-1225.
4. Barr C, Marois M, Sim I, Schmid CH, Wilsey B, Ward D, Duan N, Hays RD, Selsky J, Servadio J, Schwartz M. The PREEMPT study-evaluating smartphone-assisted n-of-1 trials in patients with chronic pain: study protocol for a randomized controlled trial. *Trials*. 2015 Dec; 16: 1-1.
5. Berg RMG, Christensen R, Ried-Larsen M. The corruption of power: On the use and abuse of a pre-trial concept. *Experimental Physiology.* 2024 Mar; 109(3):317-319.
6. Bonafiglia JT, Ross R, Gurd BJ. The application of repeated testing and monoexponential regressions to classify individual cardiorespiratory fitness responses to exercise training. *European Journal of Applied Physiology*. 2019 Apr;119(4):889-900.
7. Borenstein M, Hedges LV, Higgins JP, Rothstein HR. A basic introduction to fixed-effect and random-effects models for meta-analysis. *Research Synthesis Methods*. 2010 Apr;1(2):97-111.
8. Boulé NG, Weisnagel SJ, Lakka TA, Tremblay A, Bergman RN, Rankinen T, Leon AS, Skinner JS, Wilmore JH, Rao DC, Bouchard C. Effects of exercise training on glucose homeostasis: the HERITAGE Family Study. *Diabetes Care*. 2005 Jan; 28(1): 108-14.
9. Brennan AM, Standley RA, Yi F, Carnero EA, Sparks LM, Goodpaster BH. Individual Response Variation in the Effects of Weight Loss and Exercise on Insulin Sensitivity and Cardiometabolic Risk in Older Adults. *Frontiers in Endocrinology (Lausanne)*. 2020 Sep 10;11:632.
10. Brockwell PJ and Davis RA. Introduction to Time Series and Forecasting, 3rd ed. Ny: Springer, 2016.
11. Brooks S, Gelman A, Jones G, Meng XL, editors. Handbook of Markov Chain Monte Carlo. CRC press; 2011.

12. Chesterton P, Evans W, Wright M, Lolli L, Richardson M, Atkinson G. Influence of Lumbar Mobilizations During the Nordic Hamstring Exercise on Hamstring Measures of Knee Flexor Strength, Failure Point, and Muscle Activity: A Randomized Crossover Trial. *J Manipulative Physiol Ther*. 2021 Jan;44(1):1-13.

13. Daskalova N, Kyi E, Ouyang K, Borem A, Chen S, Park SH, Nugent N, Huang J. Self-e: Smartphone-supported guidance for customizable self-experimentation. *In Proceedings of the 2021 CHI Conference on Human Factors in Computing Systems.* 2021 May; 1-13.

14. Cox DR. Planning of experiments. New York: John Wiley & Sons, 1958.

15. Daza EJ. Causal analysis of self-tracked time series data using a counterfactual framework for n-of-1 trials. *Methods of Information in Medicine*. 2018 Feb; 57(Suppl 1): e10–e21.

16. Daza EJ, Schneider L. Model-twin randomization (motr): A Monte Carlo method for estimating the within-individual average treatment effect using wearable sensors. *arXiv preprint* arXiv:2208.00739. 2022 Aug.

17. de Brito LC, Fecchio RY, Peçanha T, Lima A, Halliwill J, Forjaz CLM. Recommendations in Post-exercise Hypotension: Concerns, Best Practices and Interpretation. *International Journal of Sports Medicine*. 2019 Aug;40(8):487-497.

18. Diabetes Prevention Program Research Group. Reduction in the incidence of type 2 diabetes with lifestyle intervention or metformin. *New England Journal of Medicine*. 2002 Feb; 346(6): 393-403.

19. Duan N, Norman D, Schmid C, Sim I, Kravitz RL. Personalized data science and personalized (N-of-1) trials: Promising paradigms for individualized health care. *Harvard Data Science Review*. 2022; 4(SI3).

20. Feng S, Mäntymäki M, Dhir A, Salmela H. How Self-tracking and the Quantified Self Promote Health and Well-being: Systematic Review. *Journal of Medical Internet Research*. 2021 Sep 21;23(9):e25171.

21. Fu J, Liu S, Du S, Ruan S, Guo X, Pan W, Sharma A, Konigorski S. Multimodal N-of-1 trials: A novel personalized healthcare design. *arXiv preprint* arXiv:2302.07547. 2023 Feb.

22. Gabler NB, Duan N, Vohra S, Kravitz RL. N-of-1 trials in the medical literature: a systematic review. *Medical Care*. 2011 Aug; 49(8): 761-8.

23. Gärtner T, Schneider J, Arnrich B, Konigorski S. Comparison of Bayesian Networks, G-estimation and linear models to estimate causal treatment effects in aggregated N-of-1 trials with carry-over effects. *BMC Medical Research Methodology*. 2023 Aug; 23(1): 191.

24. Gelman A. Prior distributions for variance parameters in hierarchical models (comment on article by Browne and Draper). *Bayesian Analysis*. 2006 Sep;1(3):515-534.

25. Gelman A, Carlin JB, Stern HS, Dunson DB, Vehtari A and Rubin DB, 2013. Bayesian Data Analysis. 3rd Edition. Chapman and Hall/CRC Press.

26. Goltz FR, Thackray AE, King JA, Dorling JL, Atkinson G, Stensel DJ. Interindividual Responses of Appetite to Acute Exercise: A Replicated Crossover Study. *Medicine and Science in Sports and Exercise*. 2018 Apr;50(4):758-768.

27. Greene WH. *Econometric Analysis*, 8th ed. NY: Pearson, 2018

28. Guolo A, Varin C. Random-effects meta-analysis: the number of studies matters. *Statistical Methods in Medical Research*. 2017 Jun;26(3):1500-1518.

29. Hecksteden A, Kraushaar J, Scharhag-Rosenberger F, Theisen D, Senn S, Meyer T. Individual response to exercise training - a statistical perspective. *Journal of Applied Physiology*. 2015 Jun 15;118(12):1450-9.

30. Hernán MA, Robins JM. Causal Inference: What If. Boca Raton: Chapman & Hall/CRC. 2020.

31. Hetherington-Rauth M, Magalhães JP, Júdice PB, Melo X, Sardinha LB. Vascular improvements in individuals with type 2 diabetes following a 1 year randomised controlled exercise intervention, irrespective of changes in cardiorespiratory fitness. *Diabetologia*. 2020 Apr;63(4):722-732.

32. Higgins JP, Thompson SG, Spiegelhalter DJ. A re-evaluation of random-effects meta-analysis. *Journal of the Royal Statistical Society Series A: Statistics in Society*. 2009 Jan;172(1):137-59.

33. Jackson D, Turner R. Power analysis for random-effects meta-analysis. *Research Synthesis Methods*. 2017 Sep;8(3):290-302.

34. Jelleyman C, Yates T, O'Donovan G, Gray LJ, King JA, Khunti K, Davies MJ. The effects of high-intensity interval training on glucose regulation and insulin resistance: a meta-analysis. *Obesity Reviews*. 2015 Nov; 16(11): 942-61.

35. Kaplan HC, Marcus GM, Yang J, Schmid CH, Schuler CL, Chang F, Dodds CM, Murphy LE, Modrow MF, Sigona K and Opipari-Arrigan L. *Understanding How to Use Single- Patient Studies to Answer Patient-Identified Research Questions.* Patient-Centered Outcomes Research Institute (PCORI), 2022.

36. Kaplan HC, Opipari-Arrigan L, Yang J, Schmid CH, Schuler CL, Saeed SA, Braly KL, Chang F, Murphy L, Dodds CM, Nuding M, Liu H, Pilley S, Stone J, Woodward G, Yokois N, Goyal A, Lee D, Yeh AM, Lee P, Gold BD, Molle-Rios Z, Zwiener RJ, Ali S, Chavannes M, Linville T, Patel A, Ayers T, Bassett M, Boyle B, Palomo P, Verstraete S, Dorsey J, Kaplan JL, Steiner SJ, Nguyen K, Burgis J and Suskind DL for the ImproveCareNow Pediatric IBD Learning Health System. Personalized research on diet in ulcerative colitis and Crohn's disease: A series of n-of-1 diet trials. *American Journal of Gastroenterology.* 2022; 117: 902-917.

37. Konigorski S, Wernicke S, Slosarek T, Zenner AM, Strelow N, Ruether DF, Henschel F, Manaswini M, Pottbäcker F, Edelman JA, Owoyele B. StudyU: A platform for designing and conducting innovative digital N-of-1 trials. *Journal of Medical Internet Research*. 2022 Jul; 24(7): e35884.

38. Kravitz RL, Schmid CH, Marois MM Wilsey B, Ward DH, Hays RD, Duan N, Wang Y, MacDonald S, Jerant A, Servadio JL, Haddad D and Sim I. Effect of mobile device–supported single-patient multi-crossover trials on treatment of chronic musculoskeletal pain: a randomized clinical trial. *JAMA Internal Medicine.* 2018; 178: 1368-1377.

39. Kravitz RL, Duan N, eds, and the DEcIDE Methods Center N-of-1 Guidance Panel (Duan N, Eslick I, Gabler NB, Kaplan HC, Kravitz RL, Larson EB, Pace WD, Schmid CH, Sim I, Vohra S). *Design and implementation of N-of-1tTrials: A user's guide.* AHRQ Publication No. 13(14)-EHC122-EF. Rockville, MD: Agency for Healthcare Research and Quality. 2014 Feb.

40. Kruschke JK. Rejecting or accepting parameter values in Bayesian estimation. *Advances in Methods and Practices in Psychological Science*. 2018 Jun;1(2):270-80.

41. Lee SY. Using Bayesian statistics in confirmatory clinical trials in the regulatory setting: a tutorial review. *BMC Medical Research Methodology*. 2024 May 7;24(1):110.

42. Liao Z, Qian M, Kronish IM, Cheung YK. Analysis of N-of-1 trials using Bayesian distributed lag model with autocorrelated errors. *Statistics in Medicine*. 2023 Jun; 42(13): 2044-60.

43. Little RJ, Rubin DB. Missing data in large data sets. *In Statistical Methods and the Improvement of Data Quality*. Academic Press. 1983 Jan; 215-243.

44. Little RJ, Rubin DB. *Statistical analysis with missing data*. 2019. John Wiley & Sons.

45. Lolli L, Gonzalez J, Shannon O, Spellanzon B, Stensel DJ, Thackray AE, Atkinson G. Understanding Treatment Response Heterogeneity Using Crossover Randomized Controlled Trials: A Primer for Exercise and Nutrition Scientists. *Int J Sport Nutr Exerc Metab*. 2025 Aug 28;35(6):465-492.

46. Magalhães JP, Hetherington-Rauth M, Júdice PB, Correia IR, Rosa GB, Henriques-Neto D, Melo X, Silva AM, Sardinha LB. Interindividual Variability in Fat Mass Response to a 1-Year Randomized Controlled Trial With Different Exercise Intensities in Type 2 Diabetes:

Implications on Glycemic Control and Vascular Function. *Frontiers in Physiology*. 2021 Sep 16;12:698971.

47. Makowski D, Ben-Shachar MS, Chen SA, Lüdecke D. Indices of effect existence and significance in the Bayesian framework. *Frontiers in Psychology*. 2019 Dec 10;10:2767.

48. Marcus GM, Modrow MF, Schmid CH, Sigona K, Nah G, Yang J, Chu TC, Joyce S, Gettabecha S, Ogomori K, Yang V, Butcher X, Hills MT, McCall D, Sciarappa K, Sim I, Pletcher MJ and Olgin JE. Individualized studies of triggers of proxysmal atrial fibrillation: the I-STOP-AFib trial. *JAMA Cardiology*. 2022; 7: 167–174.

49. Mari A, Pacini G, Murphy E, Ludvik B, Nolan JJ. A model-based method for assessing insulin sensitivity from the oral glucose tolerance test. *Diabetes Care*. 2001 Mar; 24(3): 539–548.

50. Maturana FM, Martus P, Zipfel S, NIE AM. Effectiveness of HIIE versus MICT in improving cardiometabolic risk factors in health and disease: a meta-analysis. *Medicine & Science in Sports & Exercise*. 2021 Mar; 53(3): 559-73.

51. Maturana FM, Schellhorn P, Erz G, Burgstahler C, Widmann M, Munz B, Soares RN, Murias JM, Thiel A, Nieß AM. Individual cardiovascular responsiveness to work-matched exercise within the moderate-and severe-intensity domains. *European Journal of Applied Physiology*. 2021 Jul;121(7):2039-59.

52. Maturana FM, Schellhorn P, Erz G, Burgstahler C, Widmann M, Munz B, Soares RN, Murias JM, Thiel A, Nieß AM. Individual cardiovascular responsiveness to work-matched exercise within the moderate- and severe-intensity domains. *European Journal of Applied Physiology*. 2021 Jul;121(7):2039-2059.

53. Meier D, Ensari I, Konigorski S. Designing and evaluating an online reinforcement learning agent for physical exercise recommendations in N-of-1 trials. *PMLR.* 2023 Dec; 340-352.

54. Melo X, Pinto R, Angarten V, Coimbra M, Correia D, Roque M, Reis J, Santos V, Fernhall B, Santa-Clara H. Training responsiveness of cardiorespiratory fitness and arterial stiffness following moderate-intensity continuous training and high-intensity interval training in adults with intellectual and developmental disabilities. *Journal of Intellectual Disability Research*. 2021 Dec;65(12):1058-1072.

55. Mirza RD, Punja S, Vohra S, Guyatt G. The history and development of N-of-1 trials. *Journal of the Royal Society of Medicine*. 2017 Aug; 110(8): 330-40.

56. Müller A, Konigorski S, Meißner C, Fadai T, Warren CV, Falkenberg I, Kircher T, Nestoriuc Y. Study protocol: combined N-of-1 trials to assess open-label placebo treatment for antidepressant discontinuation symptoms [FAB-study]. *BMC Psychiatry*. 2023 Oct; 23(1): 749.

57. Nikles J, Mitchell G, editors. The essential guide to N-of-1 trials in health. Dordrecht: Springer Netherlands; 2015 Jan.

58. Pan XR, Li GW, Hu YH, Wang JX, Yang WY, An ZX, Hu ZX, Xiao JZ, Cao HB, Liu PA, Jiang XG. Effects of diet and exercise in preventing NIDDM in people with impaired glucose tolerance: the Da Qing IGT and Diabetes Study. *Diabetes Care*. 1997 Apr; 20(4): 537-44.

59. Panagiotopoulou, K, Evrenoglou T, Schmid CH, Metelli S and Chaimani A. Meta-analysis models relaxing the random effects normality assumption: methodological systematic review and simulation study. *Statistical Methods in Medical Research*. 2025, in press. doi.org/10.48550/arXiv.2412.12945.

60. Piccininni M, Stensrud MJ, Shahn Z, Konigorski S. Causal inference for N-of-1 trials. *arXiv preprint* arXiv:2406.10360. 2024 Jun.

61. Pinheiro JC, Bates DM. *Mixed-effects models in S and S-PLUS.* New York, NY: Springer New York; 2000 May 5.

62. Röver C, Bender R, Dias S, Schmid CH, Schmidli H, Sturtz S, Weber S, Friede T. On weakly informative prior distributions for the heterogeneity parameter in Bayesian random-effects meta-analysis. *Research Synthesis Methods*. 2021 Jul;12(4):448-474.

63. Schmid CH. Marginal and Dynamic Regression Models for Longitudinal Data. *Statistics in Medicine*. 2001;20: 3295-3311.

64. Schmid CH, Brown EN. Bayesian hierarchical models. *Methods in Enzymology*. 2000 Jan; 321: 305-30.

65. Schmid CH, Cappelleri JC, Lau J. Bayesian methods to improve sample size approximations. *Methods in Enzymology*. 2004;383:406-27.

66. Schmid CH, Carlin BP and Welton NJ. Bayesian methods for meta-analysis. In *Handbook of Meta-Analysis*, eds. CH Schmid, T Stijnen and White IR. NY: CRC Press, 2020, 91-127.

67. Schmid CH and Duan N. Statistical design and analytic considerations for N-of-1 trials in *Design and Implementation of N-of-1 Trials: A User's Guide*, eds. Kravitz RL, Duan N. AHRQ Publication No. 13(14)-EHC122-EF. Rockville, MD: Agency for Healthcare Research and Quality; February 2014.

68. Schmid C and Yang J. Bayesian models for n-of-1 trials. *Harvard Data Science Review*, (Special Issue 3), 2022.

69. Schneider J, Gärtner T, Konigorski S. Multimodal outcomes in N-of-1 trials: Combining unsupervised learning and statistical inference. *arXiv preprint* arXiv:2309.06455. 2023 Sep.

70. Selker HP, Cohen T, D'Agostino RB, Dere WH, Ghaemi SN, Honig PK, Kaitin KI, Kaplan HC, Kravitz RL, Larholt K, McElwee NE. A useful and sustainable role for N-of-1 trials in the healthcare ecosystem. *Clinical Pharmacology & Therapeutics*. 2022 Aug; 112(2): 224-32.

71. Senarathne SG, Overstall AM, McGree JM. Bayesian adaptive N-of-1 trials for estimating population and individual treatment effects. *Statistics in Medicine*. 2020 Dec; 39(29): 4499-518.

72. Seng EK, Prieto P, Boucher G, Vives-Mestres M. Anxiety, incentives, and adherence to self-monitoring on a mobile health platform: a naturalistic longitudinal cohort study in people with headache. *Headache*. 2018 Nov; 58(10): 1541-1555.

73. Senn S. Sample size considerations for N-of-1 trials. *Statistical Methods in Medical Research*. 2019 Feb; 28(2): 372-383.

74. Senn S. The analysis of continuous data from n-of-1 trials using paired cycles: a simple tutorial. *Trials* 2024; 25: 128.

75. Shen T, Thackray AE, King JA, Alotaibi TF, Alanazi TM, Willis SA, Roberts MJ, Lolli L, Atkinson G, Stensel DJ. Are There Interindividual Responses of Cardiovascular Disease Risk Markers to Acute Exercise? A Replicate Crossover Trial. *Medicine and Science in Sports and Exercise*. 2024 Jan 1;56(1):63-72.

76. Shrestha S and Jain S. A Bayesian-bandit adaptive design for N-of-1 clinical trials. *Statistics in Medicine*. 2021; 40: 1825–1844.

77. Solomon TP. Sources of inter-individual variability in the therapeutic response of blood glucose control to exercise in type 2 diabetes: going beyond exercise dose. *Frontiers in Physiology*. 2018 Jul 13; 9: 896.

78. Spiegelhalter DJ, Freedman LS, Parmar MK. Bayesian approaches to randomized trials. *Journal of the Royal Statistical Society: Series A (Statistics in Society)*. 1994 May;157(3):357-87.

79. Stamatakis E, Ahmadi MN, Gill JMR et al. Association of wearable device-measured vigorous intermittent lifestyle physical activity with mortality. *Nature Medicine*. 2022; 28: 2521–2529.

80. Steele J, Fisher JP, Giessing J, Androulakis-Korakakis P, Wolf M, Kroeske B, Reuters R. Long-Term Time-Course of Strength Adaptation to Minimal Dose Resistance Training Through Retrospective Longitudinal Growth Modeling. *Research Quarterly for Exercise and Sport*. 2023 Dec;94(4):913-930.

81. Sylow L, Kleinert M, Richter EA, Jensen TE. Exercise-stimulated glucose uptake—regulation and implications for glycaemic control. *Nature Reviews Endocrinology*. 2017 Mar; 13(3): 133-48.

82. Sylow L, Richter EA. Current advances in our understanding of exercise as medicine in metabolic disease. *Current Opinion in Physiology*. 2019 Dec; 12: 12-9.

83. Swinton PA, Hemingway BS, Saunders B, Gualano B, Dolan E. A Statistical Framework to Interpret Individual Response to Intervention: Paving the Way for Personalized Nutrition and Exercise Prescription. *Frontiers in Nutrition*. 2018 May 28;5:41.

84. Thomas HJ, Marsh CE, Lester L, Maslen BA, Naylor LH, Green DJ. Sex differences in cardiovascular risk factor responses to resistance and endurance training in younger subjects. *American Journal of Physiology. Heart and Circulatory Physiology*. 2023 Jan 1;324(1):H67-H78.

85. Tuomilehto J, Lindström J, Eriksson JG, Valle TT, Hämäläinen H, Ilanne-Parikka P, Keinänen-Kiukaanniemi S, Laakso M, Louheranta A, Rastas M, Salminen V. Prevention of type 2 diabetes mellitus by changes in lifestyle among subjects with impaired glucose tolerance. *New England Journal of Medicine*. 2001 May; 344(18): 1343-50.

86. Van Dongen S. Prior specification in Bayesian statistics: three cautionary tales. *Journal of Theoretical Biology*. 2006 Sep 7;242(1):90-100.

87. Vetter VM, Kurth T, Konigorski S. Evaluation of easy-to-implement anti-stress interventions in a series of N-of-1 trials: study protocol of the anti-stress intervention among physicians study. *Frontiers in Psychiatry*. 2024 Aug; 15: 1420097.

88. Vetter VM, Kurth T, Konigorski S. Evaluation of two easy-to-implement digital breathing interventions in the context of daily stress levels in a series of N-of-1 trials: results from the Anti-Stress Intervention Among Physicians (ASIP) study. *medRxiv* 2025.04.09.25325513, 2025.

89. Vieira R, McDonald S, Araújo-Soares V, Sniehotta FF, Henderson R. Dynamic modelling of n-of-1 data: powerful and flexible data analytics applied to individualised studies. Health Psychol Rev. 2017 Sep;11(3):222-234.

90. Wang F, Gelfand AE. A simulation-based approach to Bayesian sample size determination for performance under a given model and for separating models. *Statistical Science*. 2002 May;1:193-208.

91. Whitney RL, Ward DH, Marois MT, Schmid CH, Sim I, Kravitz RL. Patient Perceptions of Their Own Data in mHealth Technology-Enabled N-of-1 Trials for Chronic Pain: Qualitative Study. *JMIR mHealth and uHealth*. 2018 Oct 11;6(10):e10291.

92. Williams CJ, Gurd BJ, Bonafiglia JT, Voisin S, Li Z, Harvey N, Croci I, Taylor JL, Gajanand T, Ramos JS, Fassett RG, Little JP, Francois ME, Hearon CM Jr, Sarma S, Janssen SLJE, Van Craenenbroeck EM, Beckers P, Cornelissen VA, Pattyn N, Howden EJ, Keating SE, Bye A, Stensvold D, Wisloff U, Papadimitriou I, Yan X, Bishop DJ, Eynon N, Coombes JS. A Multi-Center Comparison of O2peak Trainability Between Interval Training and Moderate Intensity Continuous Training. *Frontiers in Physiology.* 2019 Feb 5;10:19.

93. Yang J, Steingrimsson JA, Schmid CH. Sample size calculations for n-of-1 trials. *arXiv preprint* arXiv:2110.08970. 2021 Oct.

94. Yang Y, Thackray AE, Shen T, Alotaibi TF, Alanazi TM, Clifford T, Hartescu I, King JA, Roberts MJ, Willis SA, Lolli L, Atkinson G, Stensel DJ. A replicate crossover trial on the interindividual variability of sleep indices in response to acute exercise undertaken by healthy men. *Sleep*. 2025 Mar 11;48(3):zsae250.

95. Zenner AM, Böttinger E, Konigorski S. StudyMe: a new mobile app for user-centric N-of-1 Trials. *Trials*. 2022 Dec; 23(1): 1-5.
96. Zhou L, Schneider J, Arnrich B, Konigorski S. Analyzing population-level trials as N-of-1 trials: An application to gait. *Contemporary Clinical Trials Communications*. 2024 Apr; 38: 101282.
97. Zucker DR, Ruthazer R, Schmid CH, Feuer JM, Fischer PA, Kieval RI, Mogavero N, Rapoport RJ, Selker HP, Stotsky SA, Winston E and Goldenberg DL. Lessons learned combining N-of-1 trials to assess fibromyalgia therapies. *Journal of Rheumatology.* 2006; 33: 2069-2077.

## Appendix: Sample Size Calculation in N-of-1 Trials

As an example, consider calculating the required sample size to achieve sufficient statistical power for estimating a change in OGIS in the exercise study from section 2 using a design in which treatments are pairwise randomized in blocks of size two and modelled with random intercepts and random slopes. Thus, the model for the outcome $Y_{it}$ measured on individual *i* taking treatment $X_{it}$ at time $t$ is $Y_{it} = \alpha_i + \delta_i X_{it} + \epsilon_{it}$ where the residuals $\epsilon_{it}$ follow a first-order autoregressive model $\epsilon_{it} = \rho_i \epsilon_{it-1} + u_{it}$ having random errors $u_{it} \sim N(0, \sigma^2)$.

Assume we have pilot data showing that the intervention with the smallest effect reduced OGIS by an average of 14 ml/(min·m$^2$); the standard deviation of the individual treatment effects is 42 ml/(min·m$^2$); the standard deviation of the measurements in the control group is 46; and the within-individual standard deviation is 23 ml/(min·m$^2$).

The inputs to the Shiny App are shown in the Figure below. These are obtained as follows after a bit of rounding. First the variance of the random intercepts is simply the square 2116 of the between-individual standard deviation 46 in the control group. The random slopes variance 1750 is the square of the between-individual standard deviation of 42. The covariance between the intercepts and slopes is calculated by multiplying their standard deviations which are 46 and 42, respectively, by the correlation between them which we assume to be 0.5. We assume that the residual errors have the same standard deviation of 23 across individuals (note this is labelled as a standard error in the Shiny App interface), and that the repeated measurements have an exchangeable correlation structure with equal correlations between them of 0.4.

Using a minimal clinically important difference of 14 with a type 1 error of 0.05 and power of 80%, we obtain the sample size curve in below Figure which displays the tradeoff between the number of participants needed and the number of pairs of measurements per participant. Thus, the desired power and size of test can be achieved with 96 participants measured 2 times on each intervention. Slightly, but not much, smaller numbers of participants are needed if more measurements are made on each individual. Thus, only 88 participants are needed if each person is measured 5 times on each intervention and 80 participants with 7 pairs. The reason for the minimal gain in efficiency with more measurements is that the within-individual variance is much less than the between-individual variance so learning about between-individual variation through including more participants becomes a more important way to use resources.

However, the number of measurements taken per individual does have a large effect on how accurately an individual's effect can be estimated. Few measurements lead to much greater uncertainty about each person. For instance, if we choose to take 5 pairs of measurements on each individual, the standard error for the individual's treatment effect is about 16, whereas with only 2 pair, the standard error is 25. Thus, an average difference of 32 would be statistically significant for a two-tailed t-test with 5 pair of measurements, but a difference of 50 would be needed if only 2 pair were taken. We gain some power if we borrow strength from the other individuals in the study and use shrunken estimates. The standard errors decrease to 19 for 2 pair and to 14 for 5 pair. The gain is greater when an individual has fewer measurements because of the greater uncertainty when using only that individual’s data. See the end of Section 4.6 for more on this issue of learning from others.

## Sample Size Calculations for N-of-1 Trials

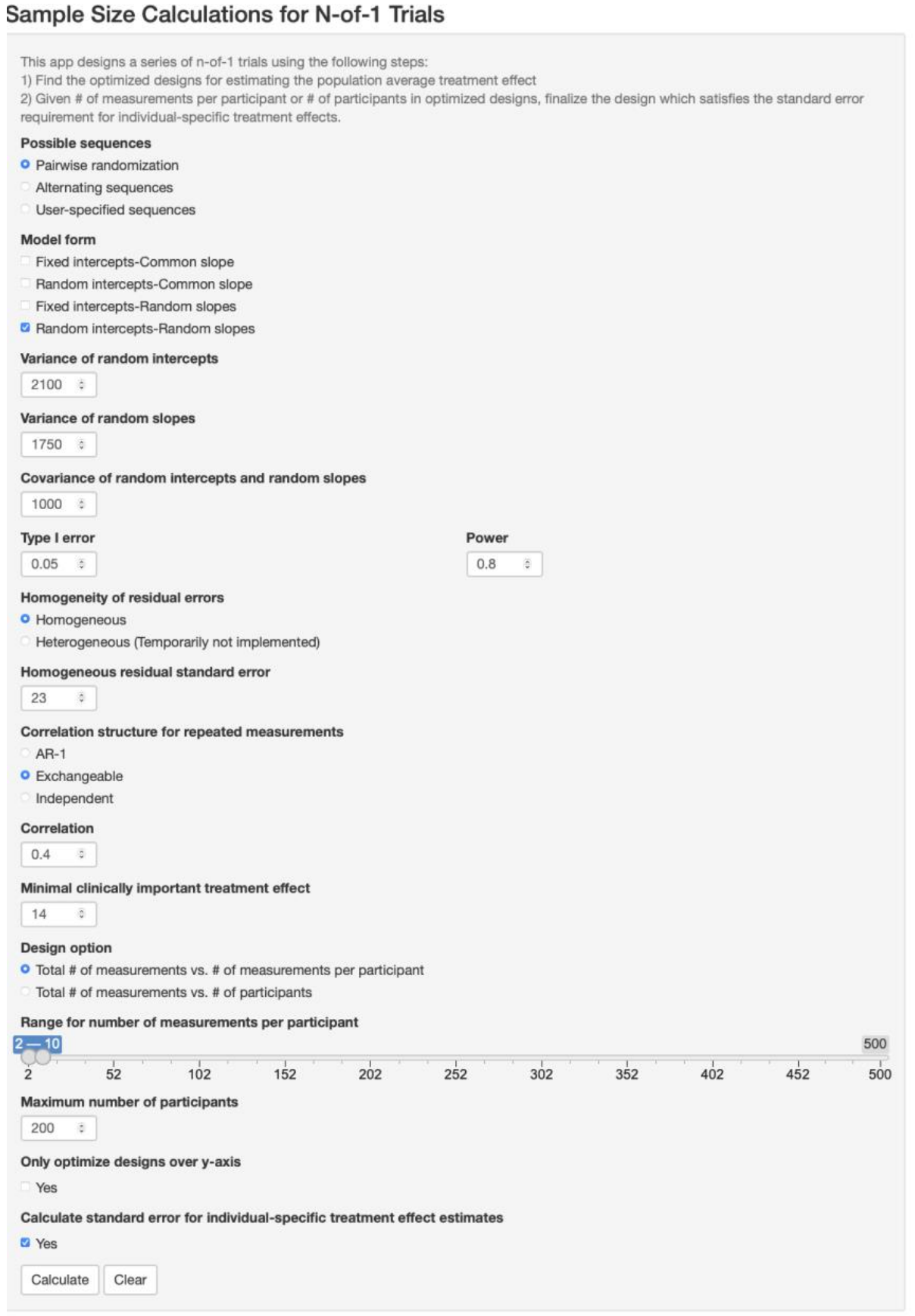

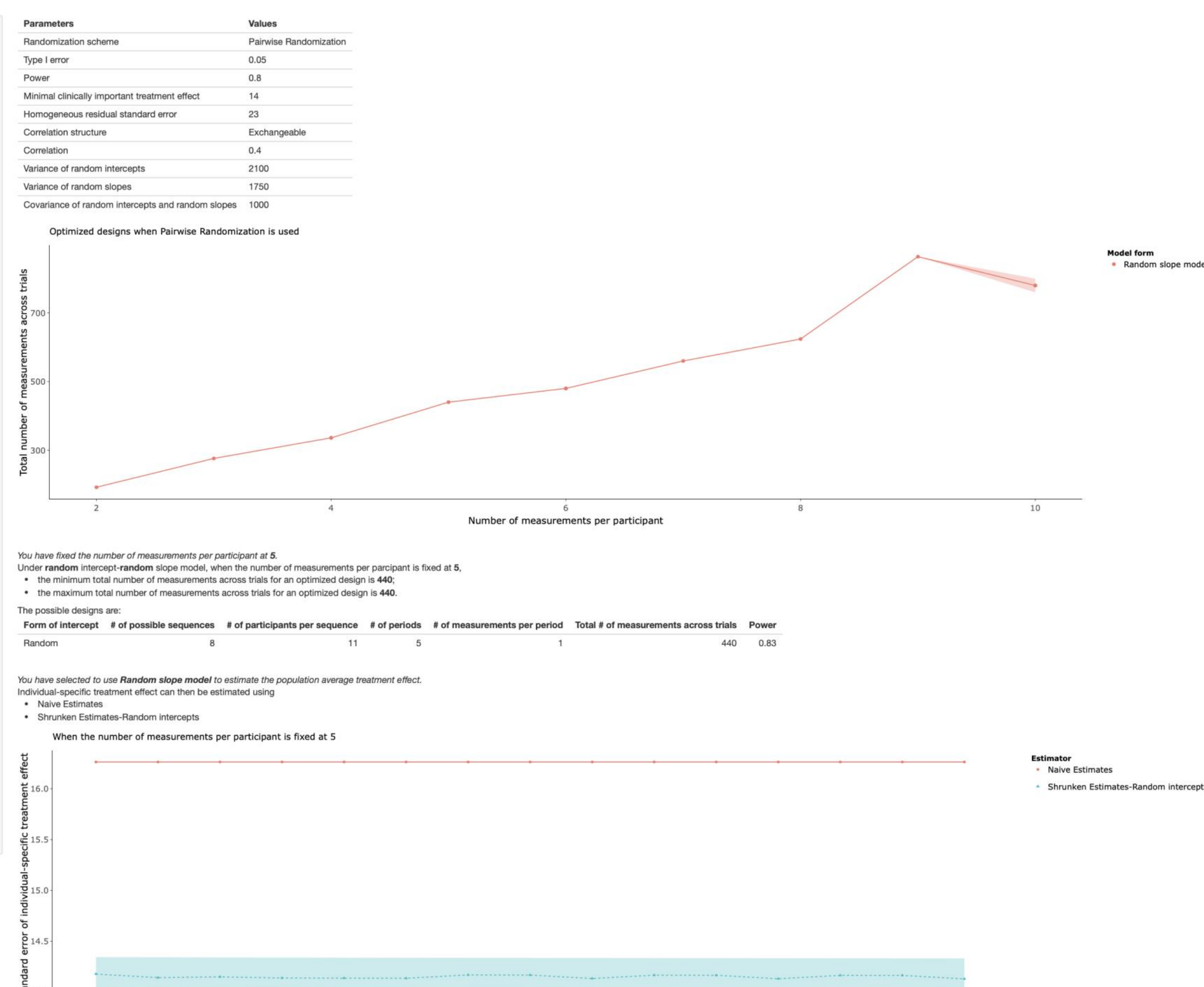

| Parameters | Values |
|---|---|
| Randomization scheme | Pairwise Randomization |
| Type I error | 0.05 |
| Power | 0.8 |
| Minimal clinically important treatment effect | 14 |
| Homogeneous residual standard error | 23 |
| Correlation structure | Exchangeable |
| Correlation | 0.4 |
| Variance of random intercepts | 2100 |
| Variance of random slopes | 1750 |
| Covariance of random intercepts and random slopes | 1000 |

*You have fixed the number of measurements per participant at **5**.*
Under **random** intercept-**random** slope model, when the number of measurements per parcipant is fixed at **5**,

- the minimum total number of measurements across trials for an optimized design is **440**;
- the maximum total number of measurements across trials for an optimized design is **440**.

The possible designs are:

| Form of intercept | # of possible sequences | # of participants per sequence | # of periods | # of measurements per period | Total # of measurements across trials | Power |
|---|---|---|---|---|---|---|
| Random | 8 | 11 | 5 | 1 | 440 | 0.83 |

*You have selected to use **Random slope model** to estimate the population average treatment effect.*
Individual-specific treatment effect can then be estimated using

- Naive Estimates
- Shrunken Estimates-Random intercepts